\documentclass[nomathfonts]{aipproc} 
\layoutstyle{6s}
\usepackage{xspace,graphicx,amsmath,amssymb,bbm}
\graphicspath{{Eps/}}
\input{alphabet}   
\input{abrege}    
\def\pth#1{\left(#1\right)}                
\def\acc#1{\left\{#1\right\}}

\def\diag{{\mathrm{diag}}}

\def\esp{{\mathrm{E}}\,}

	\def\Vect#1{\vect\bigcro{#1}}

\newsavebox{\fminibox}
\newlength{\fminilength}
\newenvironment{fminipage}[1][\linewidth]
  {\setlength{\fminilength}{#1}%-2\fboxsep-2\fboxrule}%
   \begin{lrbox}{\fminibox}\begin{minipage}{\fminilength}}
  {\end{minipage}\end{lrbox}\noindent\fbox{\usebox{\fminibox}}}

  \def\+{^\dagger}
\def\nequiv{\not\kern-.05em\equiv}
\def\egal{\kern-.5em=\kern-.5em}        % Moins d'espace autour de "="
\def\propt{\kern-.2em\propto\kern-.2em} % Idem
\def\wh#1{\widehat{#1}}                 % Sombrero !

\def\argmin{\mathop{\mathrm{arg\,min}}} % car l'indice est reparti

\def\Vect{\mathop{\text{Vect}}}

\def\intdouble{\int\kern-0.3em\int}
\def\inttriple{\int\kern-0.3em\int\kern-0.3em\int}

\def\rond#1{\overset{\kern-0.33em~_\circ}{#1}}
\def\rondit[#1]#2{\overset{\kern#1~_\circ}{#2}}

\def\babs{\begin{abstract}}             \def\eabs{\end{abstract}}
\def\barr{\begin{array}}                \def\earr{\end{array}}
\def\bcc{\begin{center}}                \def\ecc{\end{center}}

\def\bdes{\begin{description}}          \def\edes{\end{description}}
\def\bdoc{\begin{document}}             \def\edoc{\end{document}}
\def\ben{\begin{enumerate}}             \def\een{\end{enumerate}}
\def\beqn{\begin{eqnarray}}             \def\eeqn{\end{eqnarray}}
\def\beqnl#1{\beqn\label{#1}}           \def\eeqnl#1{\label{#1}\eeqn}
\def\beqnx{\begin{eqnarray*}}           \def\eeqnx{\end{eqnarray*}}
\def\bseqn{\begin{subeqnarray}}         \def\eseqn{\end{subeqnarray}}
\def\beq#1\eeq{\begin{equation}#1\end{equation}}
\def\bal#1\eal{\begin{align}#1\end{align}}
\def\balx#1\ealx{\begin{align*}#1\end{align*}}
\def\beqx{$$}                           \def\eeqx{$$}
\def\bfig{\protect\begin{figure}}       \def\efig{\protect\end{figure}}
\def\bfigx{\protect\begin{figure*}}     \def\efigx{\protect\end{figure*}}
\def\bfigt{\protect\begin{figurette}}   \def\efigt{\protect\end{figurette}}
\def\bfl{\begin{flushleft}}             \def\efl{\end{flushleft}}
\def\bfr{\begin{flushright}}            \def\efr{\end{flushright}}
\def\bit{\begin{itemize}}               \def\eit{\end{itemize}}
\def\bmi{\begin{minipage}}              \def\emi{\end{minipage}}
\def\bfmi{\begin{fminipage}}            \def\efmi{\end{fminipage}}
\def\bpic{\begin{picture}}              \def\epic{\end{picture}}
\def\bqu{\begin{quote}}                 \def\equ{\end{quote}}
\def\bqun{\begin{quotation}}            \def\equn{\end{quotation}}
\def\bsl{\begin{slide}}                 \def\esl{\end{slide}}
\def\btabb{\begin{tabbing}}             \def\etabb{\end{tabbing}}
\def\btabl{\begin{table}}               \def\etabl{\end{table}}
\def\btablx{\begin{table*}}             \def\etablx{\end{table*}}
\def\btab{\begin{tabular}} %\def\etab{\end{tabular}} Conflit avec eta gras... Elle est pas grasse, ma soeur !
\def\btabu{\begin{tabular}}             \def\etabu{\end{tabular}}
\def\btabx{\begin{tabular*}}            \def\etabx{\end{tabular*}}
\def\bbib{\begin{thebibliography}{999}} \def\ebib{\end{thebibliography}}
\def\bver{\begin{verbatim}}             \def\ever{\end{verbatim}}
\def\bca{\begin{cases}}                          \def\eca{\end{cases}}
   
\def\diag#1#2#3{\mbox{diag}(#1, #2,\ldots ,#3)}
\def\sa{\sigma _{a_{ij}}^2}
\def\iid{i.i.d.~}
\def\expf#1{\mbox{exp}\left\{#1\right\}}
\def\GExp{\Gc\Ec}
\def\Vect{\textrm{Vect}}
\def\Mbh{\widehat{\Mb}}
\def\Rbh{\widehat{\Rb}}
\def\sbh{\widehat{\sb}}
\def\ale{\alpha_{\epsilon_i}}
\def\als{\alpha_{s_j}}
\def\Re{\Rb_{\alpha_\epsilon}}
\def\Rs{\Rb_{\alpha_s}}
\def\reff#1{(\ref{#1})}
\title{Wavelet Domain Image Separation
\thanks{Presented at MaxEnt2002, the 22nd International Workshop on Bayesian 
and Maximum Entropy methods (Aug. 3-9, 2002, Moscow, Idaho, USA).  
To appear in Proceedings of American Institute of Physics}
}

\author{Ali Mohammad-Djafari}
 {
  address = {Laboratoire des Signaux et Syst\`emes,\linebreak 
  Sup\'elec, Plateau de Moulon, 91192 Gif-sur-Yvette, France},
  email = {djafari@lss.supelec.fr}
 }
 
\author{Mahieddine Ichir}
 {
  address = {Laboratoire des Signaux et Syst\`emes,\linebreak 
  Sup\'elec, Plateau de Moulon, 91192 Gif-sur-Yvette, France},
  email = {ichir@lss.supelec.fr}
 }
\begin{document}

\begin{abstract}
In this paper, we consider the problem of blind signal and 
image separation using a sparse representation of the images in the 
wavelet domain.
We consider the problem in a Bayesian estimation framework using the 
fact that the distribution of the 
wavelet coefficients of real world images can naturally be modeled by 
an exponential power probability density function. 
The Bayesian approach which has been used with success in blind source separation gives also the possibility of including any prior information 
we may have on the mixing matrix elements 
as well as on the hyperparameters (parameters of the prior laws of the 
noise and the sources). 
We consider two cases: first the case where the wavelet coefficients are 
assumed to be \iid and second the case where we model the correlation 
between the coefficients of two adjacent scales by a first order Markov 
chain. This paper only reports on the first case, the second case
results will be reported in a near future 
The estimation computations are done via a Monte Carlo Markov Chain 
(MCMC) procedure. 
Some simulations show the performances of the proposed method. 

\end{abstract}

\maketitle

\section{Introduction}
Blind source separation (BSS) is an active area of research in signal and image 
processing. Different approaches have been proposed: 
Principal component analysis (PCA) \cite{kn:Tipping99}, 
Independent factor analysis (IFA) \cite{kn:Attias,kn:Press82,kn:PS89}, 
Independent component analysis (ICA) \cite{ProcIEEE,JADE:NC,Iscas96-algebra},  
Maximum likelihood estimation 
\cite{Ziskind88,Stoica96,Wax91,InfoMaxML,ica99:lacoume,ica99:oja,ica99:macleod,ica99:bermond} 
and Bayesian estimation 
\cite{kn:Rajan97,kn:Knuth98,kn:Knuth98b,kn:LeePress,kn:Roberts98,kn:Roberts98,kn:Knuth99,Lee99a}. 
All these methods use in general independence, sparsity and diversity of 
the sources either in time or in Fourier domain. 

Wavelets, as being a powerful tool of signal processing, have been largely
used in many signal processing domains and particularly in signal denoising:
\cite{abramovich95a,abramovich98a,donoho95a,antoniadis96a,moulin98a,simoncelli99a}.
They have also been used in inverse problems:
\cite{romberg00a,wan01a,donoho92a}. 
The authors in these papers take
advantage of the properties of the wavelet coefficients
\cite{romberg00a}: locality, multi-resolution, singularity detection,
energy compaction and decorrelation. These outlined properties were said to be primary properties and
give rise to what was described to be secondary properties:
non-Gaussianity and persistency.

Zibulevsky and Pearlmutter in \cite{zibulevsky99a} considered the
problem of blind source separation within a Bayesian framework using an
\textit{over-complete} sparse representation of the sources. They have, then,
minimized an objective function assuming a known noise variance and an
empirical estimation of the sources variances.

In this paper, thanks to the unitary property of the wavelet
transform, we transport the problem of BSS to the wavelet domain and propose to use the Bayesian
estimation framework. 

According to the properties \cite{romberg00a}: decorrelation (the
wavelet coefficients of real world signals (images) tend to be \textit{approximately decorrelated}) and
non-Gaussianity (the wavelet coefficients have \textit{peaky, heavy
  tailed} marginal distributions), we propose to model the distribution of
the wavelet coefficients by a generalized exponential (GE) probability density function
(pdf). Thus, independence and sparsity which are the main hypotheses 
of all the source separation techniques are not required for the
sources themselves, but rather for their wavelet coefficients.

The Bayesian approach which has been used with success in blind 
source separation 
gives also the possibility of including any prior information we may have 
on the mixing matrix elements as well as on the hyperparameters (parameters 
of the prior laws of the noise and the sources) of the problem. 

In this work, we make use of the fast wavelet transform developed by 
Mallat \cite{mal99a} 
to have a non-redundant multi-scale representation. 
This paper is organized as follows: 
In section 2, we first present the general 
source separation problem using notation which can be used either in 
the 1D, 2D or the m-D case. Then, we write the same problem in the wavelet domain 
and explicit our hypotheses about the prior distributions of the noise and 
wavelet coefficients. 
In section 3, we present the Bayesian approach and give the main expressions 
of the prior and posterior probability density functions. 
In section 4, first we give the basics of the MCMC algorithm and then apply 
it to our case. 
In section 5, we present a few simulation results to show the performances 
of the proposed method and give some comparison with other known and classical 
approaches. 
Finally, in section 6, we present our conclusions and perspectives. 

\section{Problem Formulation}
\label{Problem}
Blind image separation consists of estimating sources from a set of
their linear mixtures. The observations consist of $m$ images 
$\{X_i, i=1,\ldots,m\}$ which
are instantaneous linear mixtures of $n$ unknown sources 
$\{S_j, j=1,\ldots,n\}$, possibly
corrupted by additive noise $\{\xi_i, i=1,\ldots,m\}$: 
\beq
X=A S+\xi
\eeq
where $A_{(m\times n)}$ is the mixing matrix. 
To be able to consider 1D, 2D or even m-D signals, we assume that $X_i$, 
$S_j$ and $\xi_i$ contain each $T$ samples representing either $T$ samples 
of time series or $T$ pixels of an image or, more generally, $T$ voxels 
of an m-D signal. Thus, $S$ is a $(n\times T)$ matrix and $X$ and $\xi$ 
are $(m\times T)$ matrices.

The blind source separation problem is to estimate both the mixing matrix 
$\Ab$ and the sources $S$ from the data $X$ and some assumptions about noise 
distribution and some prior knowledge of sources distributions. 
Different approaches have been proposed: 
Principal component analysis (PCA) \cite{hyv01a,dja99a} mainly assumes the problem without noise and Gaussian distribution for sources, 
Independent component analysis (ICA) \cite{hyv01a,lee98a} and 
Maximum likelihood estimation \cite{dja99a} assume again the problem without 
noise but different non-Gaussian distributions for sources, Factor analysis 
(FA) methods take account of the noise, but assume Gaussian priors both for 
the noise and the sources. 

The Bayesian approach is a generalization of
FA with the possibility of any non-Gaussian priors for noise and sources 
as well as the possibility of accounting for any prior knowledge on the elements 
of the mixing matrix and the hyperparameters of the problem. 
In addition, it allows us to jointly estimate the sources $S$, the mixing 
matrix $\Ab$ and even the hyperparameters $\thetab$ of the problem through 
the posterior:
\beq
p(S,\Ab,\thetab | X) \propto p(X | S,\Ab,\thetab)~ p(S | \thetab)~ 
p(\Ab | \thetab)~ p(\thetab)
\eeq
We have used this approach before with different priors $p(S | \thetab)$ 
such as Gaussian \cite{Djafari00a} and mixture of Gaussians \cite{Snoussi00a,Snoussi00b}. 
We also used this approach in multi-spectral image separation in astronomy 
for separating the cosmological microwave background (CMB) from other 
cosmological microwave activities \cite{Snoussi01a,Snoussi01b,Snoussi01c,Snoussi01d,Snoussi01e,Snoussi01f}.

In this paper, we are going to use the same Bayesian approach, but doing 
the separation taking the advantage of the independence and diversity
properties of the wavelet domain 
coefficients of the sources. 
Noting by the vector $\sb$ the $T$ samples of one of the sources, by $\Hb$ 
the discrete wavelet transform matrix, and by $\wb$ the complete
wavelet coefficients of the 1-D signal we have 
\beq
\sb =\Hb\wb
\eeq
Now, using the fact that the complete discrete wavelet transform is a linear 
and unitary operator $(\Hb^t\Hb=\Hb\Hb^t=\Ib)$, the problem of source 
separation can be easily transported to the wavelet domain and written as:
\beq
W_x=A W_s+W_{\xi}
\eeq
The main advantage of using this last equation in place of the original 
source separation problem is that we can more easily assign simple prior 
laws for $W_s$ than for $S$ itself. For example, when $S$ contains 
discontinuity or non-stationary, still its wavelet coefficients distribution 
can be modeled by a simple generalized exponential (GE) probability density 
function (pdf) while it is harder to model appropriately signal samples 
distribution by a simple pdf. Indeed, it has been reported by many authors 
that the distribution of the wavelet coefficients of real world images are 
well modeled by a GE pdf:
\beq
p(w | \alpha ,\beta )=\GExp(\alpha,\beta)=
\frac{\beta}{2\alpha \Gamma (1/\beta )} 
\expf{-|w/\alpha |^\beta}
\label{expo}
\eeq
Note that $\beta=1$ gives an exponential pdf and $\beta=2$ corresponds to 
a Gaussian pdf.  
We are going to use this prior probability law in our Bayesian estimation 
framework. 

This is shown in the following figures. Figure \reff{original} shows
two images (Lena and the cameraman) which we will use later in our
simulations. Figure \reff{hist_original} shows their respective histograms while
Figure \reff{wavelet_original} shows their wavelet coefficients and Figure
\reff{wave_hist} shows the corresponding histograms of their wavelet
coefficients. We can remark that even if the histograms of the image
pixels are very different, the corresponding wavelet coefficients are
similar and can be modeled easily by GE pdf, with different $\alpha
\rm{~and~} \beta$. For a given signal or image, these two parameters
can be estimated using either the Maximum Likelihood (ML) method:
$$(\hat{\alpha},\hat{\beta})=\argmin_{(\alpha,\beta)}\bigg(n\ln \alpha
+n\ln \frac{\Gamma(\frac{1}{\beta})}{\beta}+\frac{1}{\alpha^\beta}
\sum_{i=1}^{n}|x_i|^\beta\bigg)$$
or a moments based method by noting that the moments of the GE pdf are
given by:
\begin{displaymath}
\esp{\{x^n\}}=
\left\{
\begin{array}{cl}
 \frac{\Gamma(\frac{n+1}{\beta})}{\Gamma(\frac{1}{\beta})}
\alpha^n & \textrm{if $n$ is even}\\
\\
0 & \textrm{if $n$ is odd}
\end{array}
\right.
\end{displaymath}

\bfig[!htb]
\begin{tabular}{@{}c@{~}c@{}}
\includegraphics[width=65mm,height=65mm]{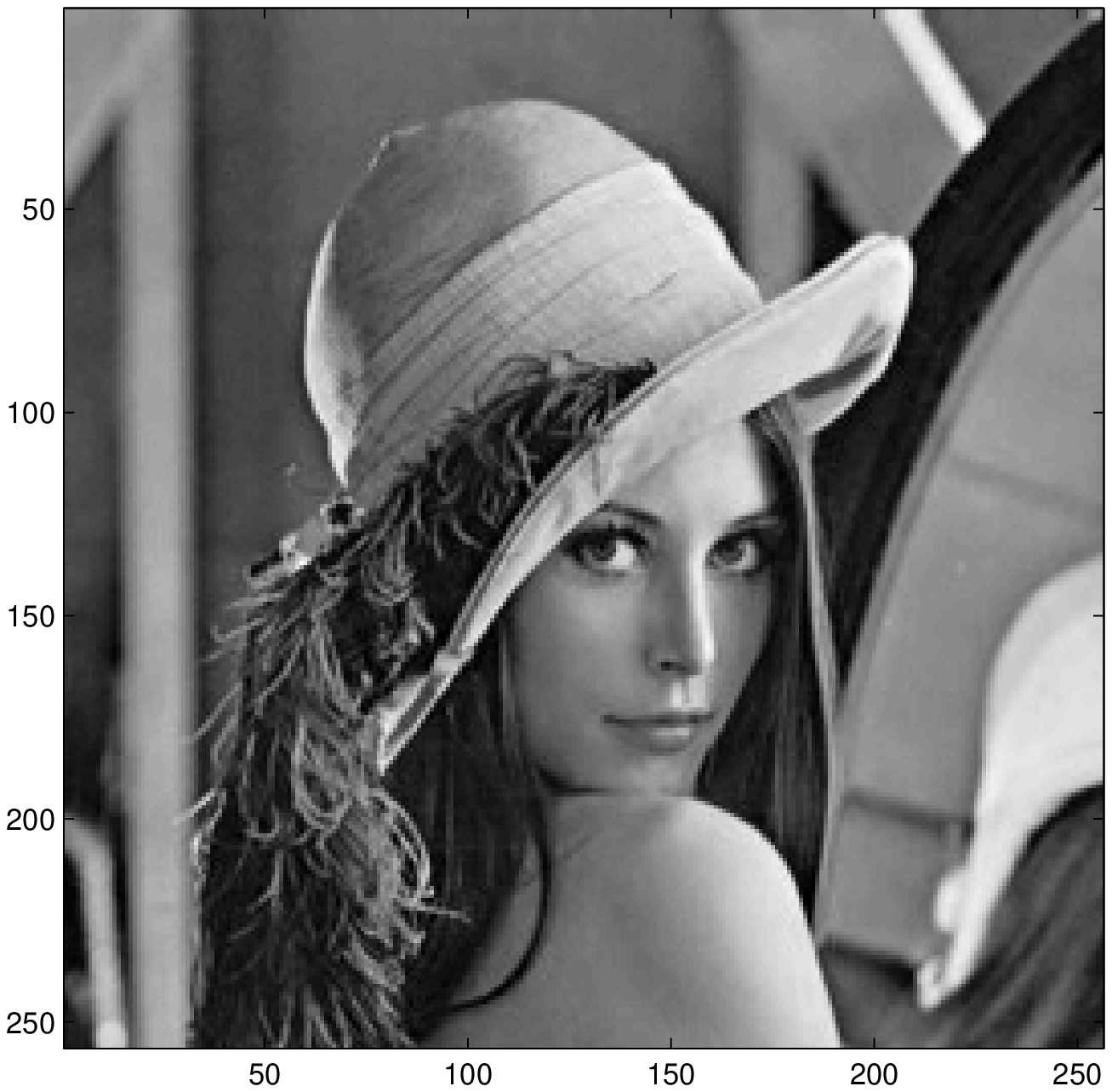} &
\includegraphics[width=65mm,height=65mm]{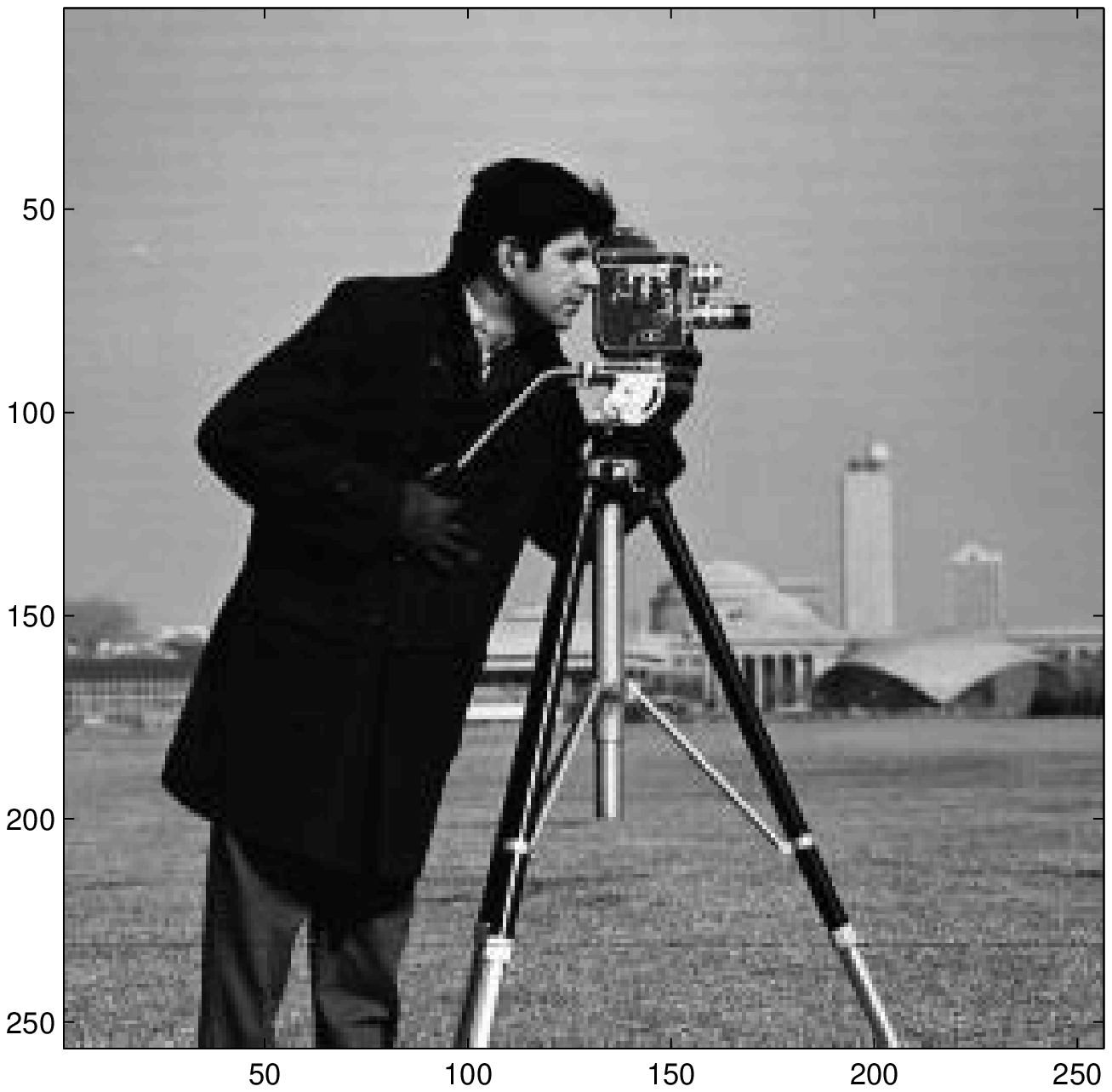}
\end{tabular}
\caption{Lena and the cameraman images}
\label{original}
\efig

\bfig[!htb]
\begin{tabular}{@{}c@{~}c@{}}
\includegraphics[width=65mm,height=40mm]{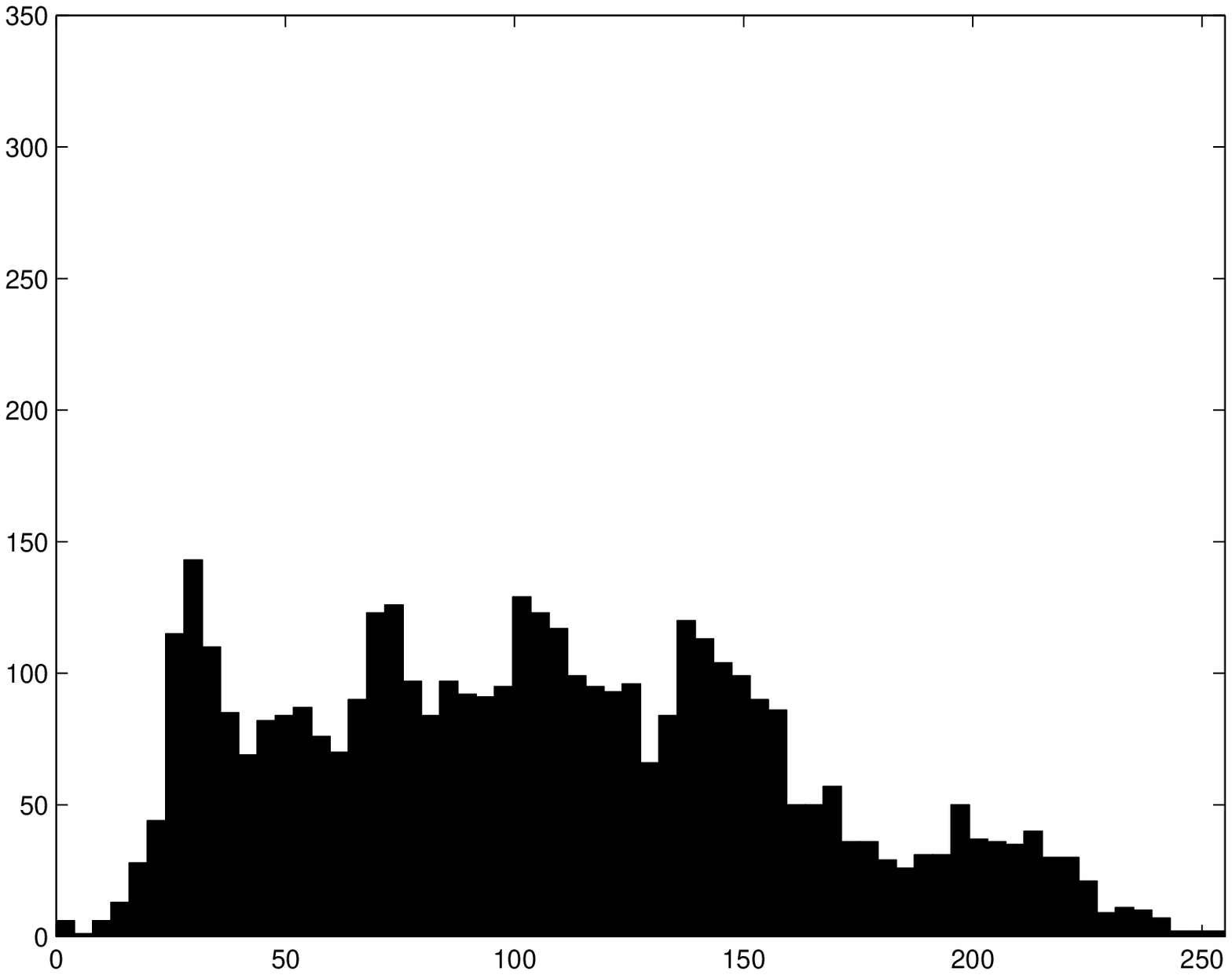} &
\includegraphics[width=65mm,height=40mm]{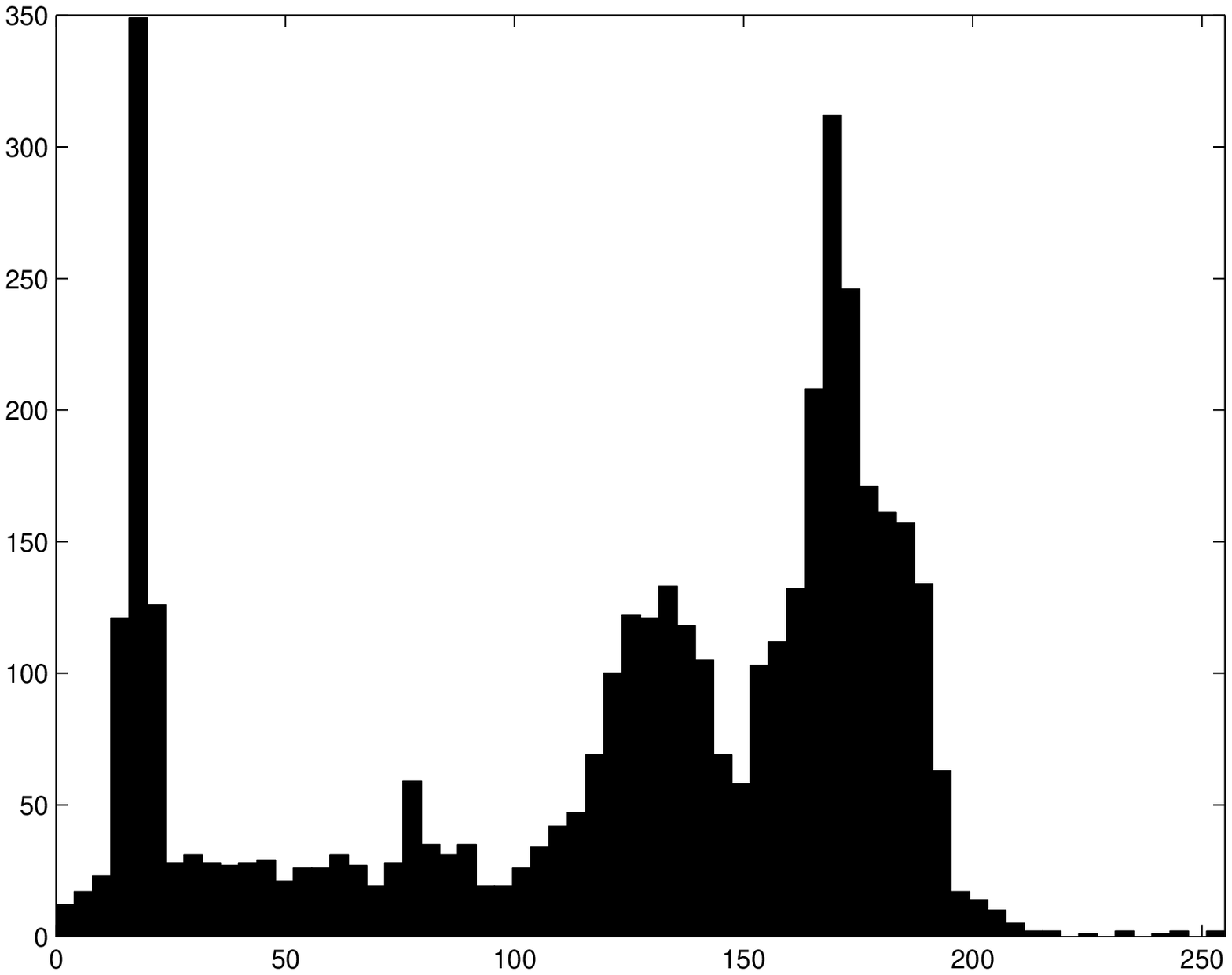}
\end{tabular}
\caption{Histograms of Lena and the cameraman images}
\label{hist_original}
\efig

\bfig[!htb]
\begin{tabular}{@{}c@{~}c@{}}  
\includegraphics[width=65mm,height=65mm]{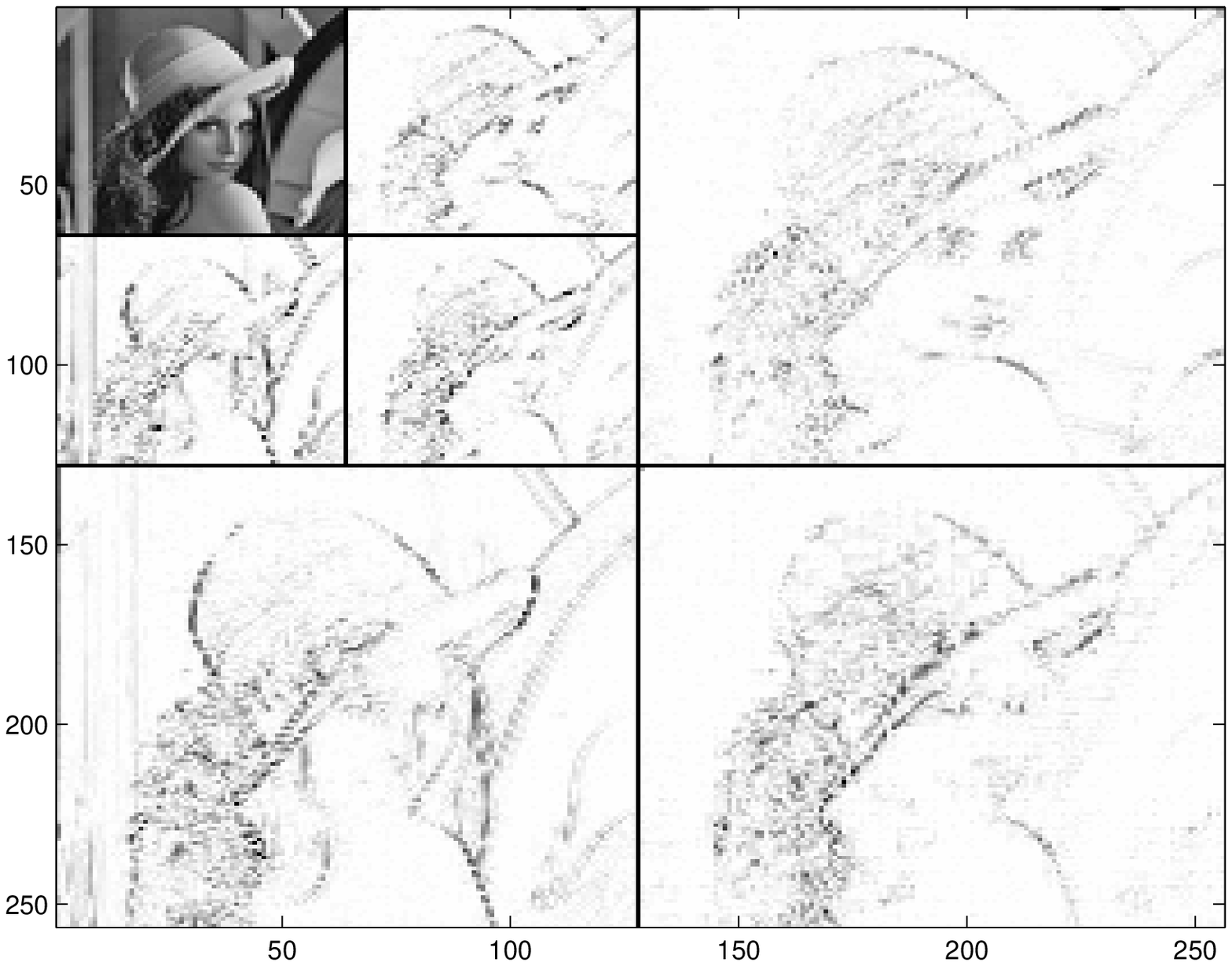} &
\includegraphics[width=65mm,height=65mm]{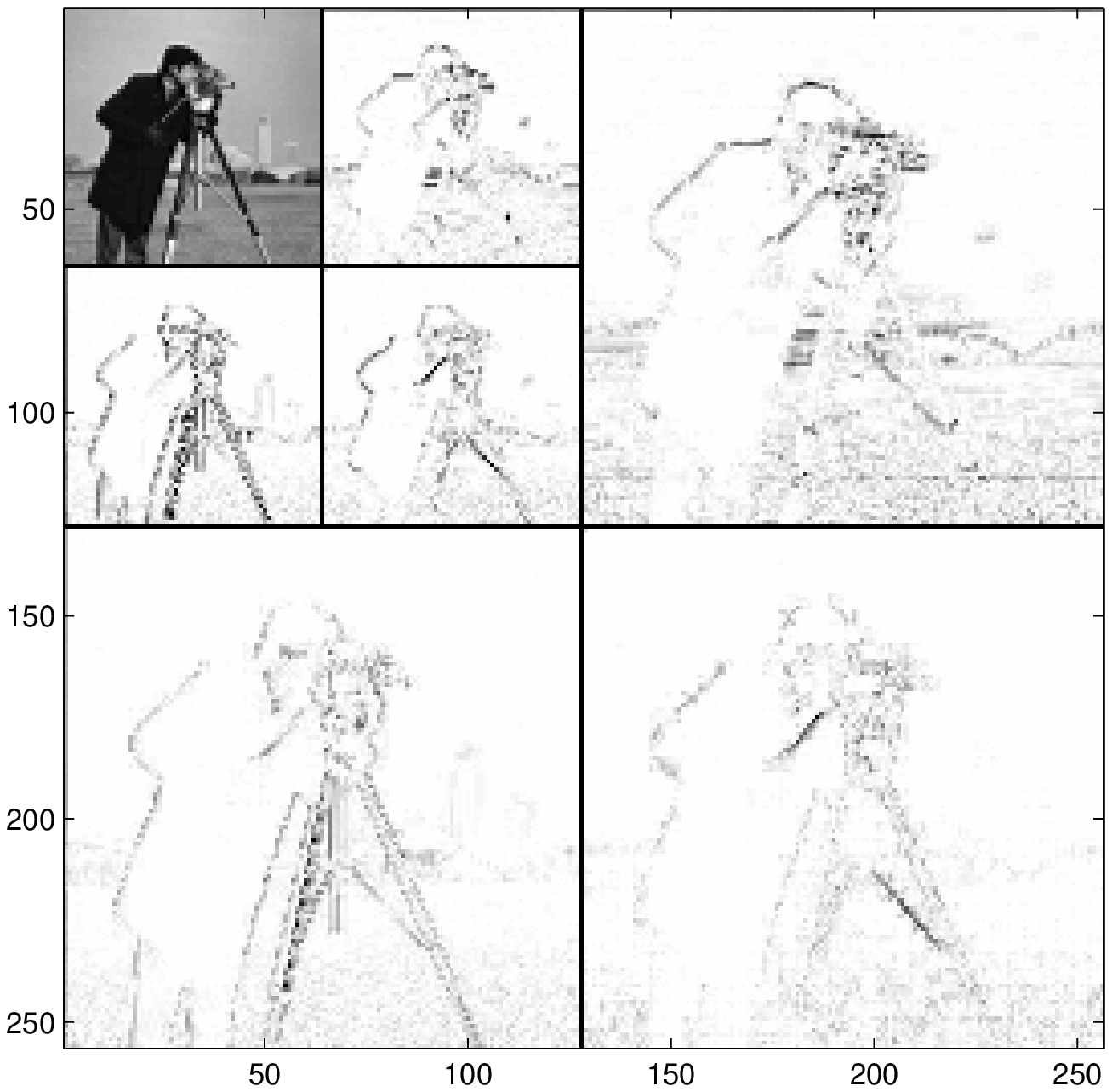} \\
\end{tabular}
\caption{Wavelet coefficients of Lena and the cameraman images}
\label{wavelet_original}
\efig

\bfig[!htb]
\begin{tabular}{@{}c@{~}c@{}}  %  sep des 2 images : 1 & 2
\includegraphics[width=65mm,height=40mm]{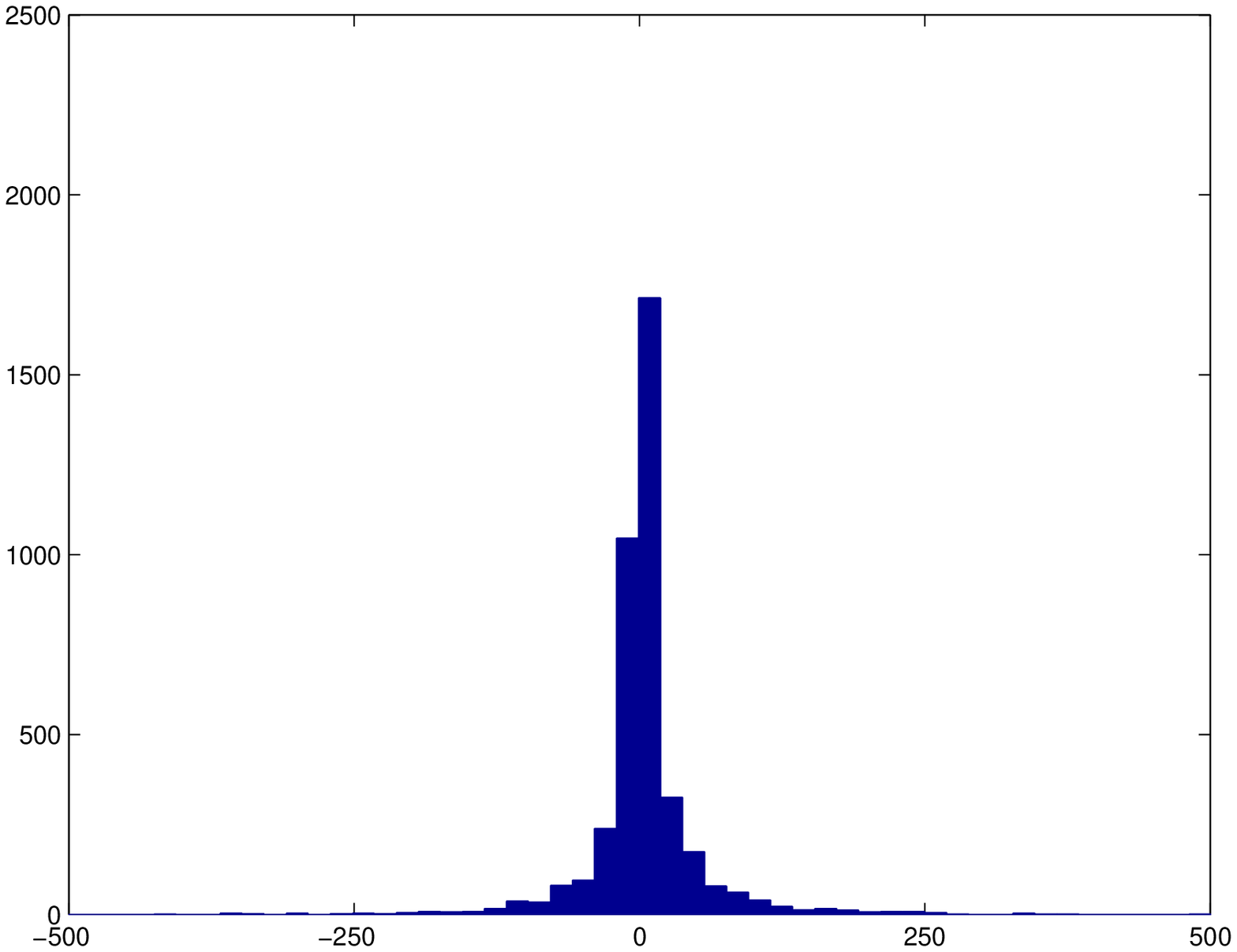} &
\includegraphics[width=65mm,height=40mm]{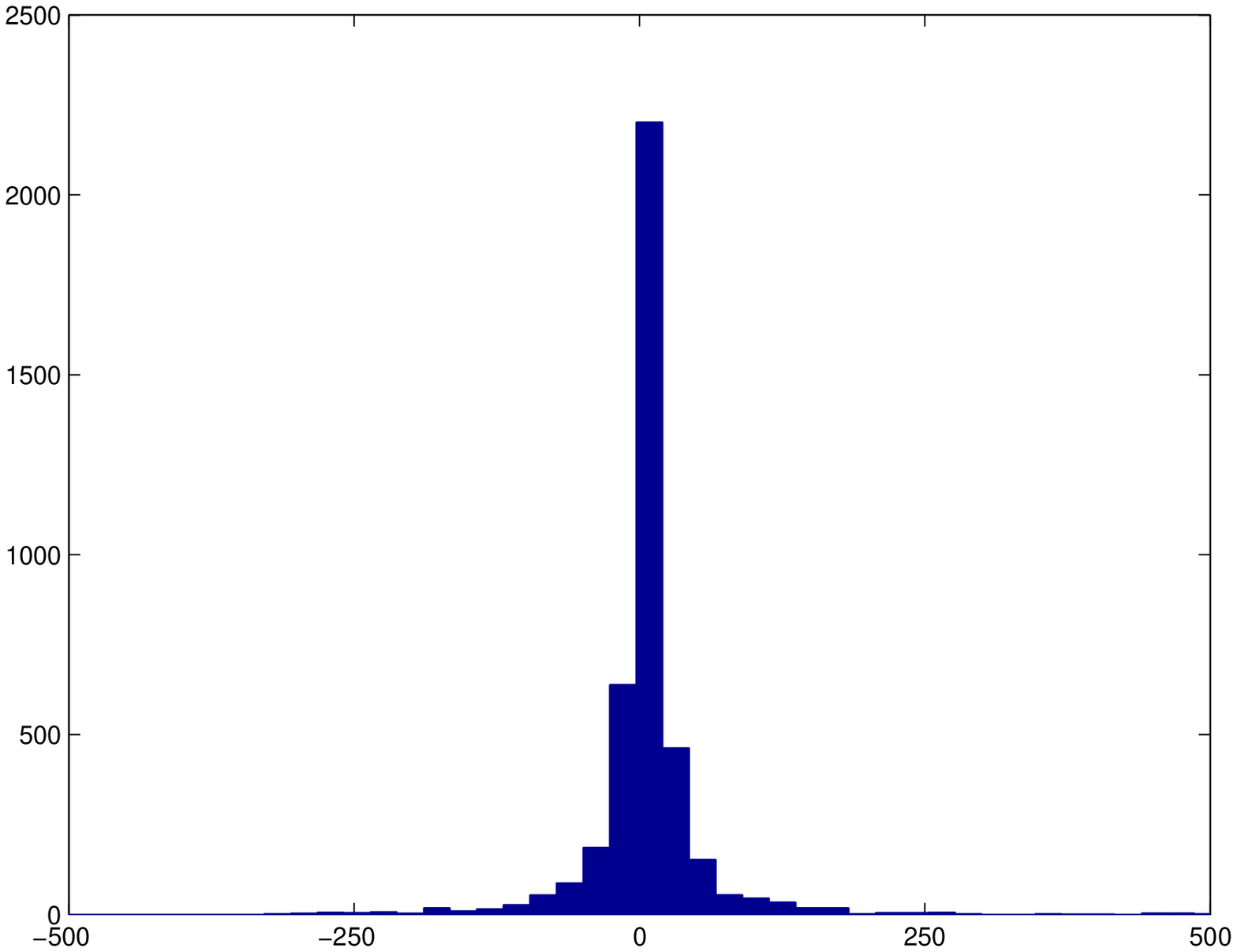} \\
\end{tabular}
\caption{Histograms of the wavelet coefficients of Lena and the cameraman images}
\label{wave_hist}
\efig

\section{Bayesian Formulation}
\label{Bayesian Formulation}
In a first step, we assume that the sources and the noise wavelet coefficients 
are \iid. Thus, to simplify the notation, we denote, respectively, by $\xb(k)$, 
$\sb(k)$ and $\xib(k)$ the vectors containing the wavelet coefficients of 
the data, the sources and the noise for a given index $k$. Thus, we have 
$\xb(k)=\Ab\sb(k)+\xib(k)$. Hereafter, we omit the index $k$ and note it 
only when needed. 
To proceed with the Bayesian approach, we have to assign the prior laws. 
In the following we assume:

\bit
\item 
The noise wavelet coefficients $\xib$ are assumed independent and 
$p(\xi_i)=\GExp(\ale,\beta)$. Then 
\beq
p(\xb | \Ab,\sb,\{\ale ,\beta\} )=\prod_{i=1}^{m}
\left(\frac{\beta}{2\ale \Gamma (1/\beta)}\right)
\expf{-\sum_{i=1}^{m}\big(|x_i-[\Ab \sb]_i|/\ale\big)^\beta}
\eeq

\item 
The wavelet coefficients $\sb$ of the sources are also assumed independent 
and $p(s_j)=\GExp(\als,\beta_s)$. Then
\beq
p(\sb | \acc{\als,\beta_s})=\prod_{j=1}^{n}
\left(\frac{\beta_s}{2\als \Gamma (1/\beta_s )}\right)
\expf{-\sum_{j=1}^{n}\big(|s_j|/\als \big)^{\beta_s}}
\eeq

\item 
The elements $a_{ij}$ of the mixing matrix $\Ab$ are assumed 
\iid and Gaussian with mean values $\mu_{ij}$ and variances $\sigma_{ij}^2$:
\beq
p(a_{ij})=(2\pi\sa)^{-1/2}\expf{-\frac{1}{{2\sigma_{a_{ij}}^2}}(a_{ij}-\mu_{ij})^2}
\eeq
Therefore, we may note by 
\begin{eqnarray}
p(\Ab | \Mb,\Rb_a) 
&=& (2\pi)^{-{mn}/2} |\Rb_a|^{-1/2}\nonumber \\ 
& & \exp\acc{-\frac{1}{2} 
\big(\Vect(\Ab-\Mb)\big)^t\Rb_a^{-1}
\big(\Vect(\Ab-\Mb)\big)}
\end{eqnarray}
where $\Mb=\acc{\mu_{ij}}$, $\Vect(\Mb)$ means a vector containing 
the elements of the matrix $\Mb$ and 
\[
\Rb_{a~(mn\times mn)}=
\diag{{\sigma _{a_{11}}^2}}{{\sigma _{a_{12}}^2}}{{\sigma _{a_{mn}}^2}}
\]

\item
All the hyperparameters 
$(\frac{1}{\ale ^\beta }, \frac{1}{\als^{\beta _s}})$ are assumed 
independent and assigned standard Gamma prior distributions
$p(x)=\Gc(2,1)$, where:
\beq
{\Gc}(x|a,b)= \frac{x^{a-1}}{b^a \Gamma(a)} \exp (-\frac{x}{b})
\eeq
\eit
The joint \apost law of the sources coefficients $\sb$, the mixing matrix $\Ab$ 
and the hyperparameters $\thetab$ is then given by:
\beq
p(\sb,\Ab,\thetab | \xb)\propto 
p(\xb | \sb,\Ab,\thetab)~p(\sb | \thetab)~p(\Ab | \thetab)~p(\thetab)
\label{apost}
\eeq
where we noted all the hyperparameters $\pth{{1\over\ale^{\beta}},{1\over\als^{\beta_s}}}$ 
by $\thetab$. 

The conditional \apost laws of $\sb, \Ab$ and $\thetab$ are then given by :
\begin{eqnarray}
p(\sb | \xb, \Ab, \thetab )
& \propto & \prod_{i=1}^{m}\left(\frac{\beta}{2\ale \Gamma (1/\beta)}\right)~\prod_{j=1}^{n}
\left(\frac{\beta_s}{2\als \Gamma (1/\beta_s )}\right) \nonumber \\
& & \expf{-\sum_{i=1}^{m}\big({|x_i-[\Ab \sb]_i|/\ale}\big)^\beta-\sum_{j=1}^{n}\big({|s_j|/\als}\big)^{\beta_s}}
\label{pds}
\end{eqnarray}

\begin{eqnarray}
p(\Ab | \xb, \sb, \thetab )&\propto &
\prod_{i=1}^{m}\left(\frac{\beta}{2\ale \Gamma
    (1/\beta)}\right)~(2\pi)^{-{mn}/2}|\Rb_a|^{-1/2} \nonumber \\
& &\expf{-\sum_{i=1}^{m}\big({|x_i-[\Ab \sb]_i|/\ale}\big)^\beta} \nonumber \\
& &\expf{-\frac{1}{2} \big(\Vect(\Ab-\Mb)\big)^t\Rb_a^{-1} \big(\Vect(\Ab-\Mb)\big)}
\label{pda}
\end{eqnarray}

\begin{eqnarray}
p\bigg(\theta_i={1\over\ale^\beta} | \xb, \sb, \Ab \bigg)&\propto & \left(\frac{\beta}{2\Gamma (1/\beta)}\right)^{K} 
\pth{\frac{1}{\ale^\beta}}^{{K\over\beta}+1} \nonumber\\
& & \expf{{1\over\ale^\beta}\pth{-\sum_{k=1}^{K}{|x_i(k)-[\Ab \sb (k)]_i|}^\beta+1}}
\label{pdthe}
\end{eqnarray}
\begin{eqnarray}
p\bigg(\theta_j={1\over\als^{\beta_s}} | \xb, \sb, \Ab \bigg)&\propto & \left(\frac{\beta_s}{2\Gamma (1/\beta_s)}\right)^{K} \pth{1\over \als^{\beta_s}}^{{K\over\beta_s}+1} \nonumber \\
& & \expf{{1\over\als^{\beta_s}}\pth{-\sum_{k=1}^{K}{|s_j(k)|}^{\beta_s}+1}}
\label{pdthej}
\end{eqnarray}

\section{MCMC Implementation}
\label{MCMC}
Once the expression of the joint \apost law $p(\sb,\Ab,\thetab | \xb)$ 
of all the unknowns has been derived, we can use it to infer them.  
However, in general, the computation of the normalization factor needs 
a huge dimensional integration. When the MAP estimation is chosen, this 
normalization factor is not needed, but it is formally needed 
for other estimation rules such as the posterior mean. 
The MCMC algorithms are then the basic tools to generate samples from 
the posterior law. The main idea is to generate successively the samples 
from the posterior laws  
$\sb^{(k)} \sim p(\sb | \Ab^{(k)},\thetab^{(k)},\xb)$, 
$\Ab^{(k)} \sim p(\Ab | \sb^{(k)},\thetab^{(k)},\xb)$ 
and 
$\thetab^{(k)} \sim p(\thetab | \sb^{(k)},\Ab^{(k)},\xb)$ 
and then estimate their expected values by averaging these samples. 

We use the Hasting-Metropolis algorithm combined to a Gibbs sampler 
to obtain an ergodic chain, and then approximate the ensemble expectation 
of any quantity $Z$ by its empirical mean:
\[
\esp{(Z)} \approx \frac{1}{(N-T+1)}\sum_{t=T}^{N}h(Z^{(t)})
\]
where $\{Z^{(t)}\}$ are samples from $p(z | .)$. 

Noting that, when $\beta=2$ and $\beta_s=2$, the posterior laws for the sources 
and for the elements of the mixing matrix are 
Gaussian, we can use these Gaussian as the trial (or instrumental) pdf. 
Thus, to simplify the presentation 
of the proposed algorithm, we give here the expressions of these Gaussian 
posterior laws: 
\bit
\item The trial posterior pdf of the sources is Gaussian 
$g(\sb | \thetab,\xb)=\Nc(\sbh, \Rbh_s)$ with 
\beq
\sbh=2\Rbh_s \Ab^t\Re^{-1} \xb
\label{sbh}
\eeq
and
\beq
\Rbh_s={1\over2}(\Ab^t\Re^{-1} \Ab+\Rs^{-1})^{-1}
\label{Rbhs}
\eeq
where
\begin{eqnarray*}
\Rb_{\alpha_s~(n\times n)}&=&\diag{{\alpha_{s_1}^2}}{{\alpha_{s_2}^2}}{{\alpha_{s_n}^2}} \\
\Rb_{\alpha_\epsilon~(m\times m)}&=&\diag{{\alpha_{\epsilon_1}^2}}{{\alpha_{\epsilon_2}^2}}{{\alpha_{\epsilon_m}^2}}
\end{eqnarray*}

\item The trial posterior pdf of the mixing matrix elements is Gaussian 
$g(\Vect(\Ab) | \thetab,\xb)=\Nc(\Vect(\Mbh), \Rbh_a)$ with 
\beq
\Vect(\Mbh)=\Rbh_a \left(2\Vect(\sb \xb^t \Re^{-1})+\Rb_a^{-1}\Vect(\Mb)
\right)
\label{Mbh}
\eeq
and
\beq
\Rbh_a=\left(2\left(\Eb^t\Re^{-1}\Eb\right).*\Cb+\Rb_a^{-1}\right)^{-1}
\label{Rbha}
\eeq
where
\begin{eqnarray*}
\Eb_{(m\times mn)}&=&\textrm{blockdiag}\left([1,\ldots,1]_{(n\times 1)},m\right)\\
\Cb_{(mn\times mn)}&=&\textrm{blockdiag}\left(ss^t_{(n\times n)},m\right)
\end{eqnarray*}
where blockdiag$(\Mb,m)$ stands for a $m$ block-diagonal matrix with
matrix $\Mb$ as the block elements, and $\Ab.*\Bb$ stands for a
point-wise multiplication of two matrices, i.e. $\Cb=\Ab.*\Bb$ means $\Cb_{ij}=\Ab_{ij}\Bb_{ij}$.

\eit
The proposed MCMC algorithm is then the following:
\begin{itemize}
\item Initialize $\sb,\Ab,\thetab $ ~to~ $\sb^0,\Ab^0,\thetab ^0$ and 
repeat the following steps until convergence
\item Sampling $\sb(k)$, for $k=1\ldots K$:
\[
\zb\leadsto g(\zb | \thetab,\xb)=\Nc(\sbh,\Rbh_s)
\]
where $\sbh$ and $\Rbh_s$ are given, respectively by eq.~(\ref{sbh}) 
and eq.~(\ref{Rbhs}) and 
\[
\sb^{(t+1)}(k)=\left\{
\barr{cll}
\zb & \textrm{with probability} & \rho \\
\sb^{(t)}(k) & \textrm{with probability} & 1-\rho
\earr
\right.
\]
with 
\[
\rho=\textrm{min}\left(
1,\frac{p(\zb | \xb(k),\Ab,\thetab)}{p(\sb^{(t)}(k) | \xb(k),\Ab,\thetab)}
  \bigg/ 
  \frac{g(\zb)}{g(\sb^{(t)}(k))}
\right)
\]
where $p(\zb | \xb(k),\Ab,\thetab)$ is given by eq.(\ref{pds}).
\item Sampling $\Ab$:
\[
\zb\leadsto g(\zb| \thetab,\xb)=\Nc(\textrm{Vect}(\Mbh),\Rbh_a)
\]
where $\Mbh$ and $\Rbh_a$ are given, respectively by eq.~(\ref{Mbh})  
and eq.~(\ref{Rbha}) and 
\[
\Ab^{(t+1)}=\left\{
\barr{cll}
\textrm{Mat}(\zb) & \textrm{with probability} & \rho \\
\Ab^{(t)} & \textrm{with probability} & 1-\rho
\earr
\right.
\]
with
\[
\rho=\textrm{min}\left(
1,\frac{p(\zb | \xb,\sb,\thetab)}{p(\Vect(\Ab^{(t)}) | \xb,\sb,\thetab)}
\bigg/ 
\frac{g(\zb)}{g(\Vect(\Ab^{(t)}))}
\right)
\]
where $p(\zb | \xb(k),\Ab,\thetab)$ is given by eq.~(\ref{pda}).
\item Sampling $\theta_i=\frac{1}{\ale^\beta}$, for $i=1\ldots m$:
\[
\thetab^{(t+1)}\leadsto \Gc(a,b)
\]
with
\[
a=\frac{K}{\beta}+2
\mbox{~~~and~~~}
b=\left(
\sum_{k=1}^{K}|x_i(k)-[\Ab\sb(k)]_i|^{\beta}+1\right)^{-1}
\]
\item Sampling $\theta_j=\frac{1}{\als^{\beta_s}}$, for $j=1\ldots n$:
\[
\thetab_j^{(t+1)}\leadsto \Gc(a,b)
\]
with
\[
a=\frac{K}{\beta_s}+2
\mbox{~~~and~~~}
b=\Big(\sum_{k=1}^{K}|s_j(k)|^{\beta_s }+1\Big)^{-1}.
\]
\end{itemize} 

\section{Simulation results}
\label{Simulation}
To illustrate the performances of the proposed method, we consider two
cases: a favorable case where we have 2 unknown sources with 3
measured data, and a more difficult case where we have only two
measured data. In the first case, we consider $64 \times 64$ pixel
images of the two images of Figure
\reff{original} with the following rectangular mixing matrix: 

\[
A=\left[
\begin{array}{cc}
  0.8211 &   0.4053 \\
  0.3769 &   0.7997 \\
  0.4287 &   0.4428 
\end{array}
\right]
\]
to generate the mixed images and added a white Gaussian noise of zero
mean to obtain the data with a SNR $=30$dB, where SNR
is defined as being the ratio of the mixed signal energy to that of
the noise in dB: $\textrm{SNR}_{\textrm{dB}} = 10\log_{10} \left({\|\xb\|^2\over\|\epsilonb\|^2}\right)$. Figure \reff{mixed_rect} shows the mixed images obtained.

\bfig[!htb]
\btabu{@{}c@{~}c@{~}c@{}}
\includegraphics[width=45mm,height=45mm]{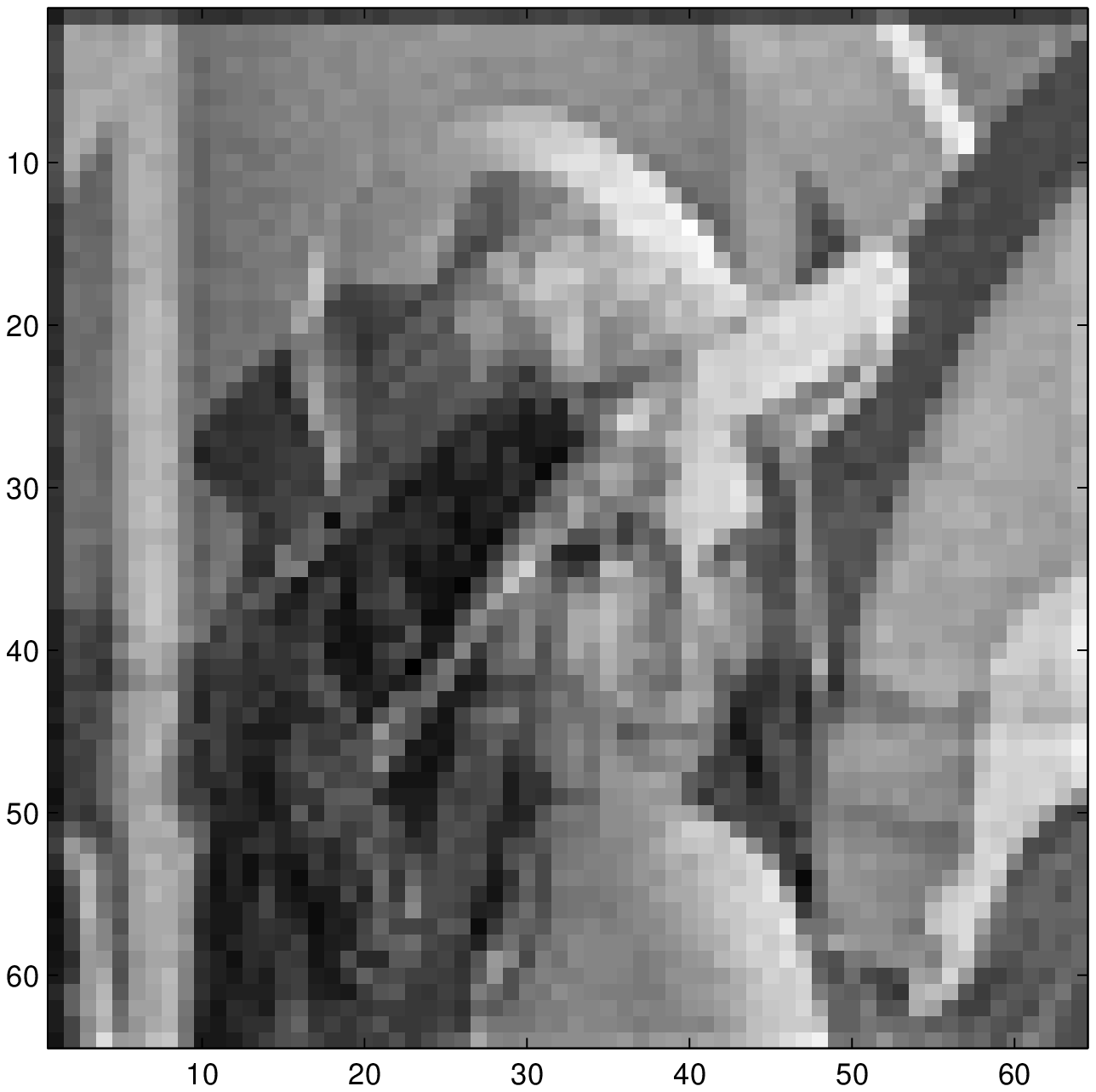} &
\includegraphics[width=45mm,height=45mm]{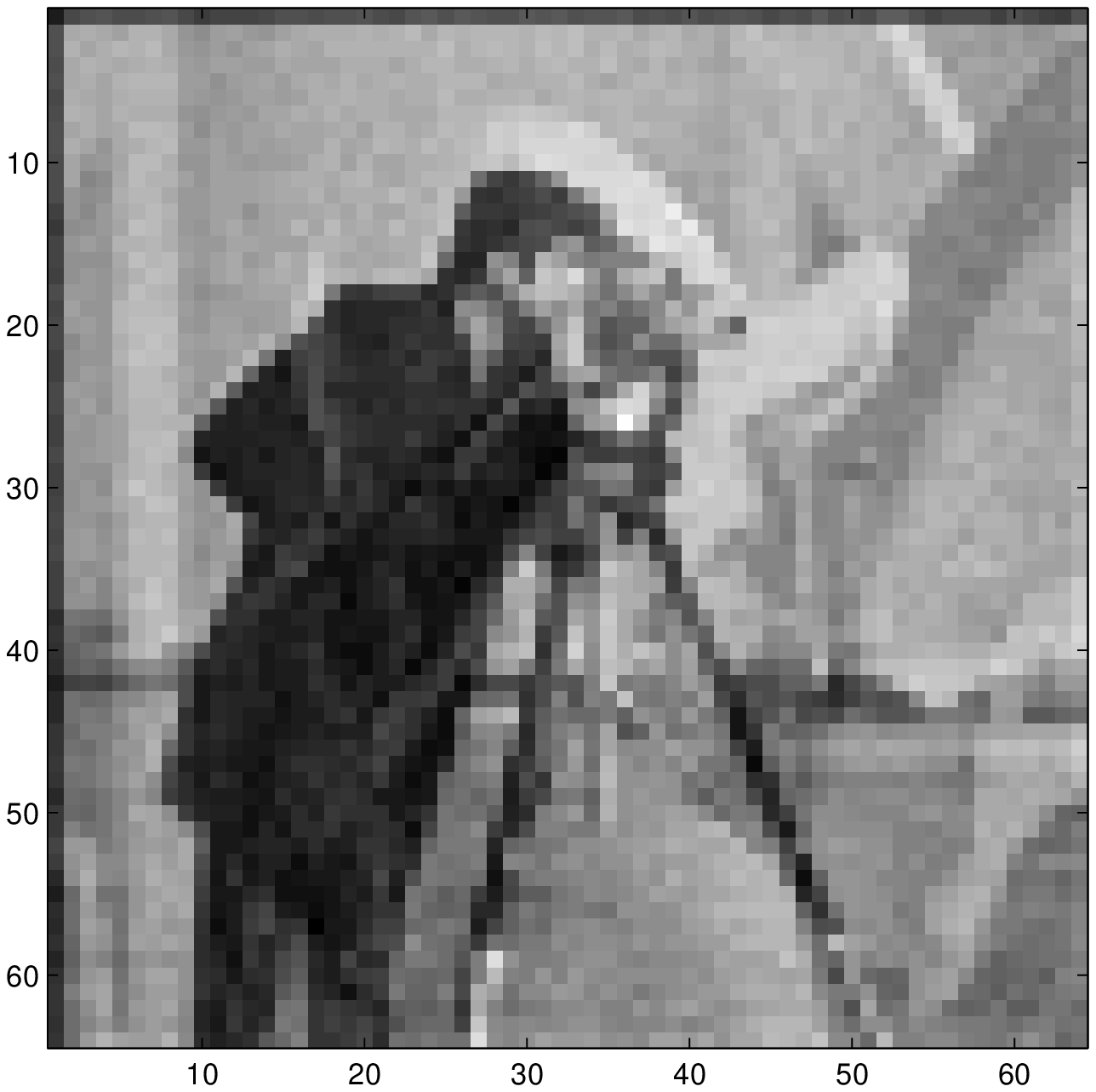} &
\includegraphics[width=45mm,height=45mm]{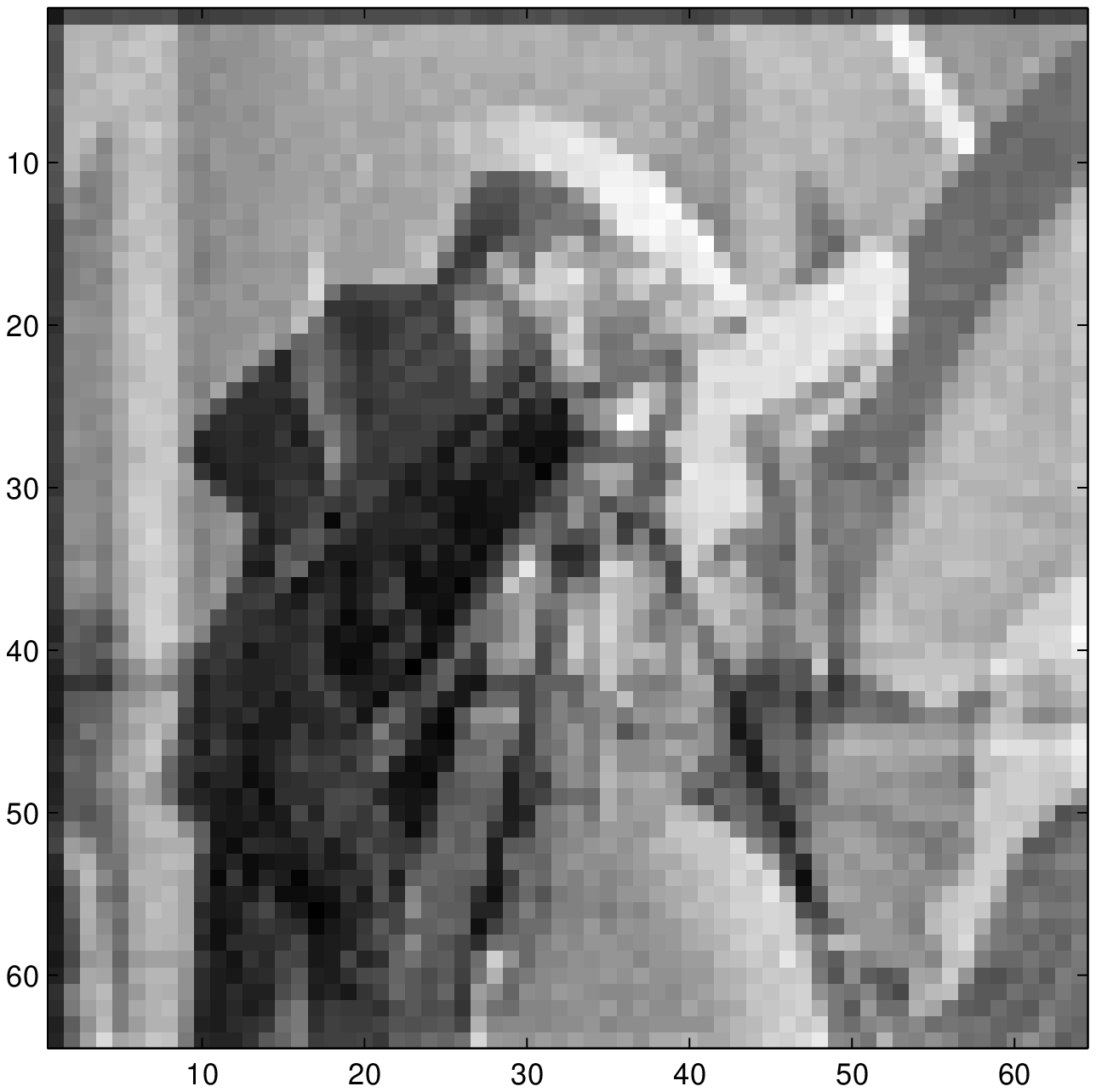}
\etabu
\caption{The mixed images in the rectangular case}
\label{mixed_rect}
\efig

We applied the proposed method directly on the mixed images 
where we assumed noise to be \iid and original images to be independent and 
Gaussian. Then, we accounted for the local correlation between
neighboring pixels through a Markovian modeling of the original
images. Finally, we applied the method in the wavelet domain. Figure \reff{estimated_rect} shows the
separated images obtained for each case.
  
\bfig[!hbt]
\btabu{@{}c@{~}c@{~}c@{}}
(a) & (b) & (c) \\
\includegraphics[width=45mm,height=45mm]{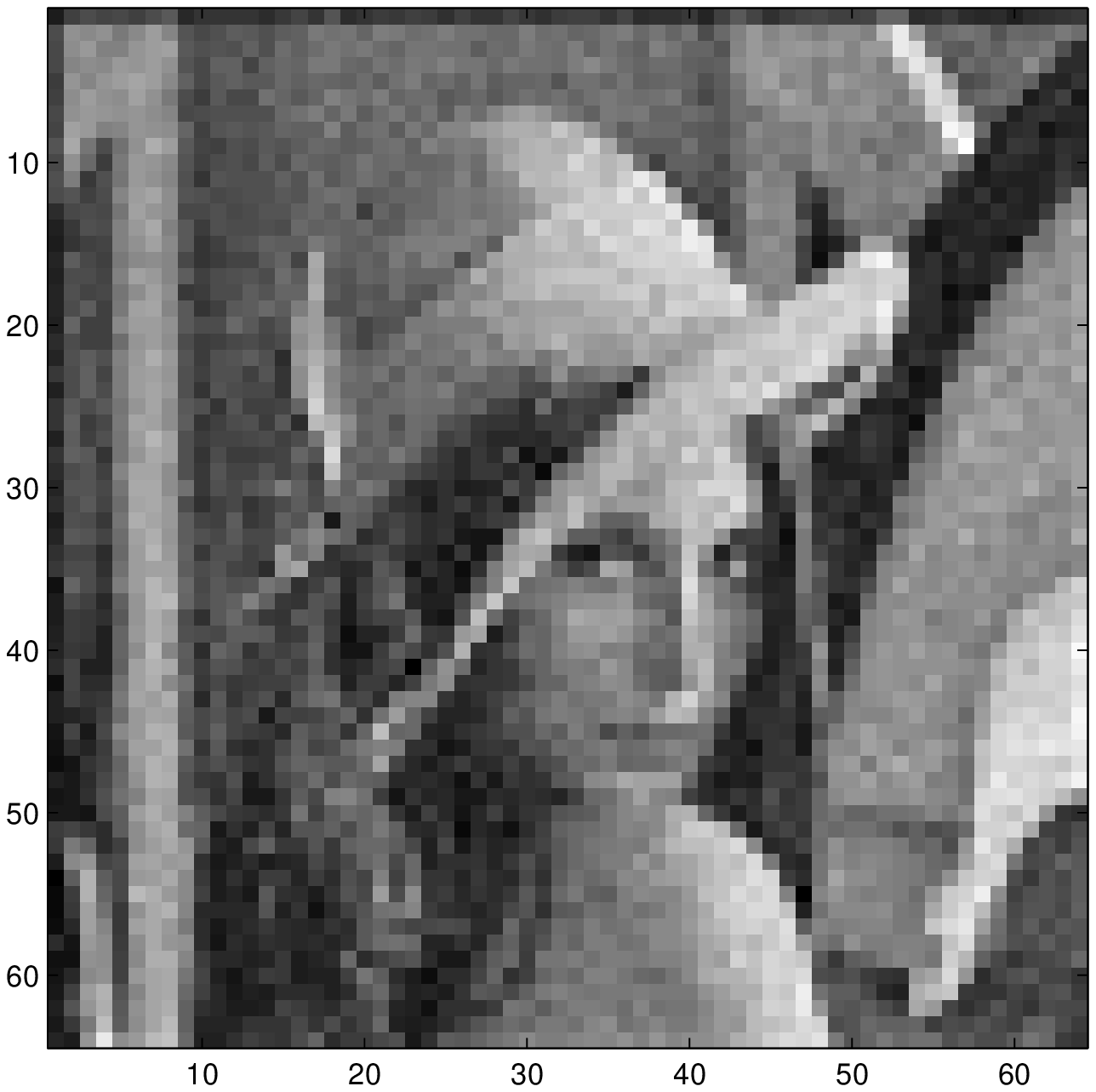} &
\includegraphics[width=45mm,height=45mm]{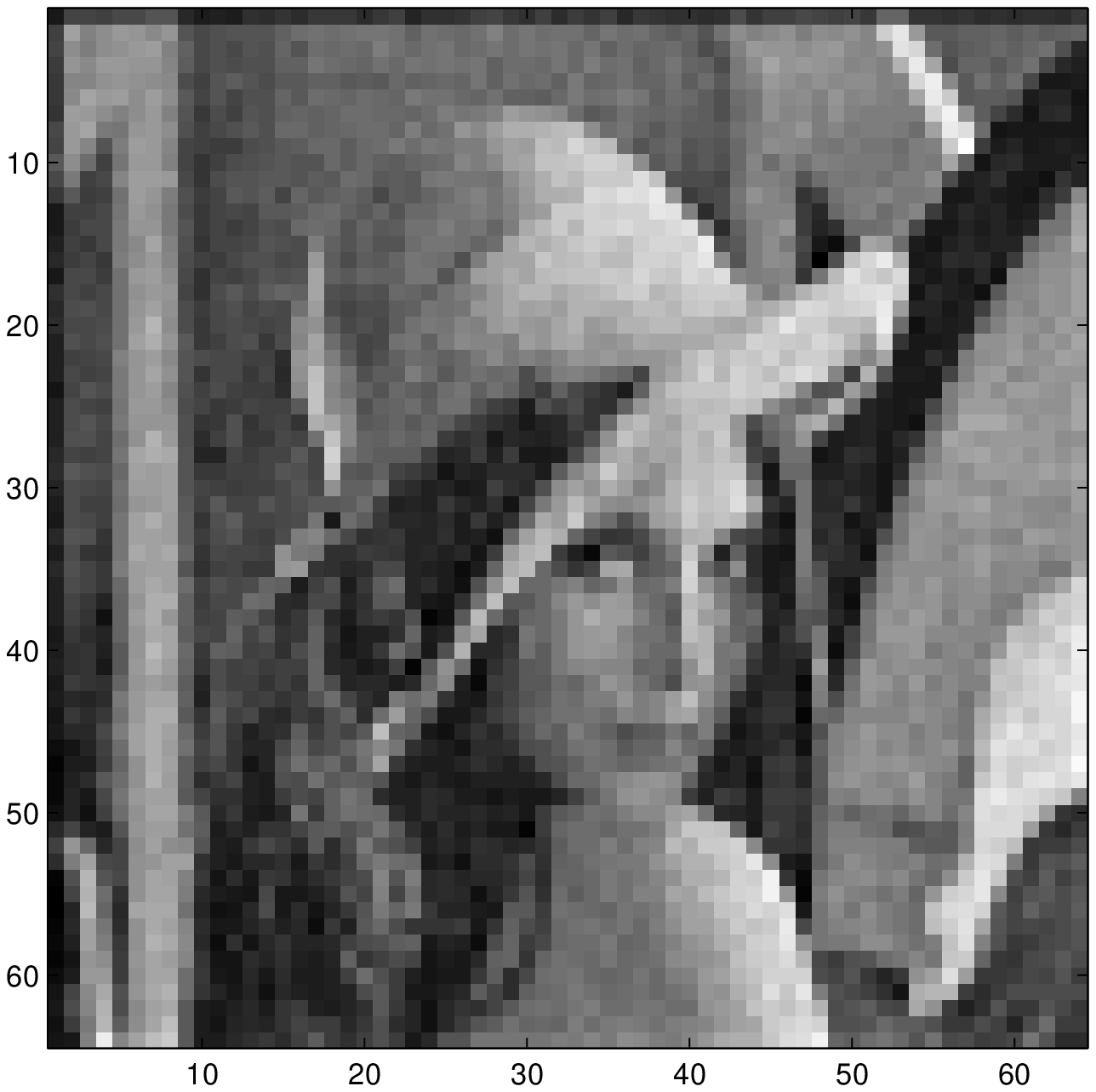} &
\includegraphics[width=45mm,height=45mm]{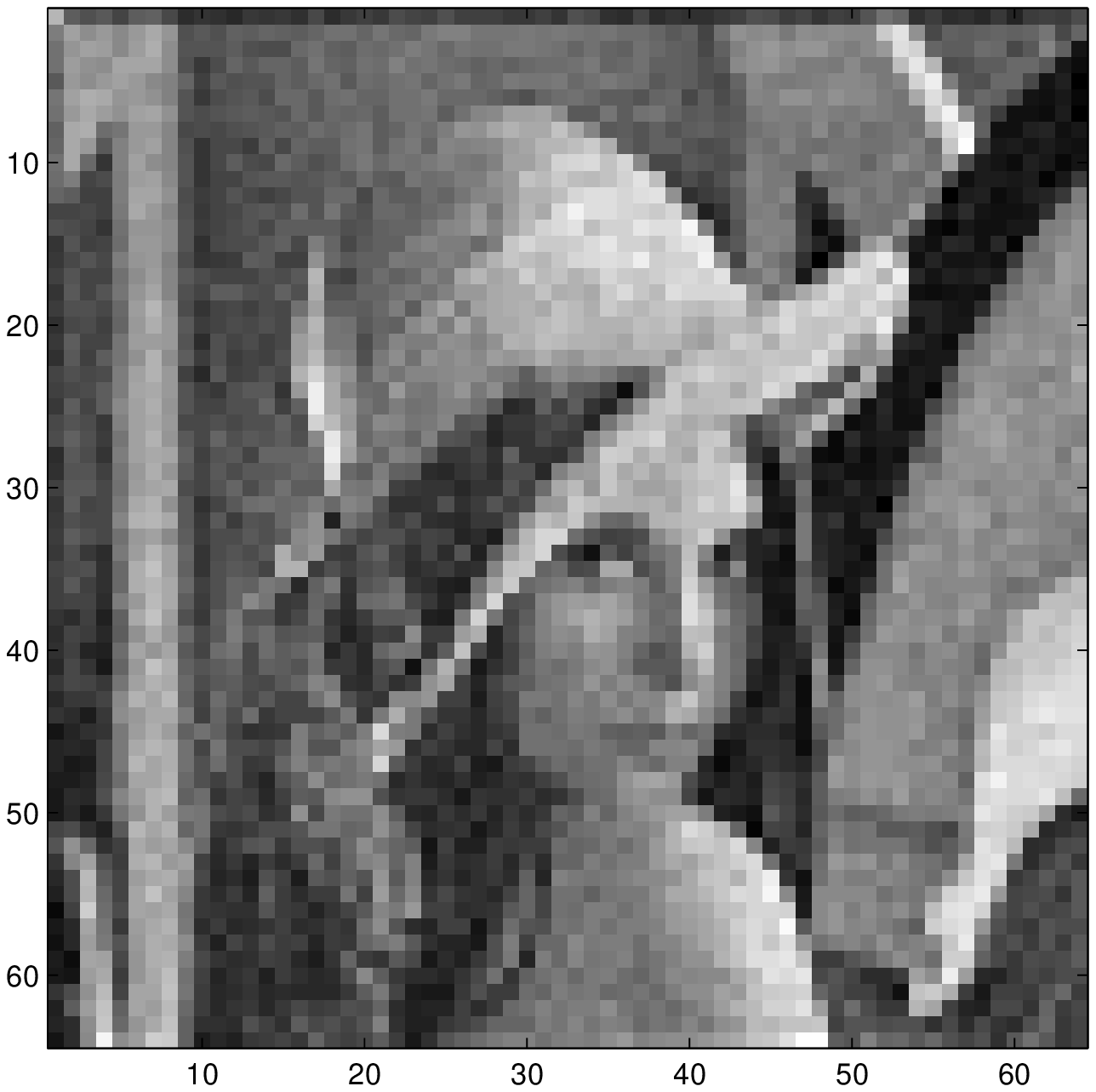} \\
PSNR = $0.0910$ & PSNR = $0.0651$ & PSNR = $0.1144$ \\
\includegraphics[width=45mm,height=45mm]{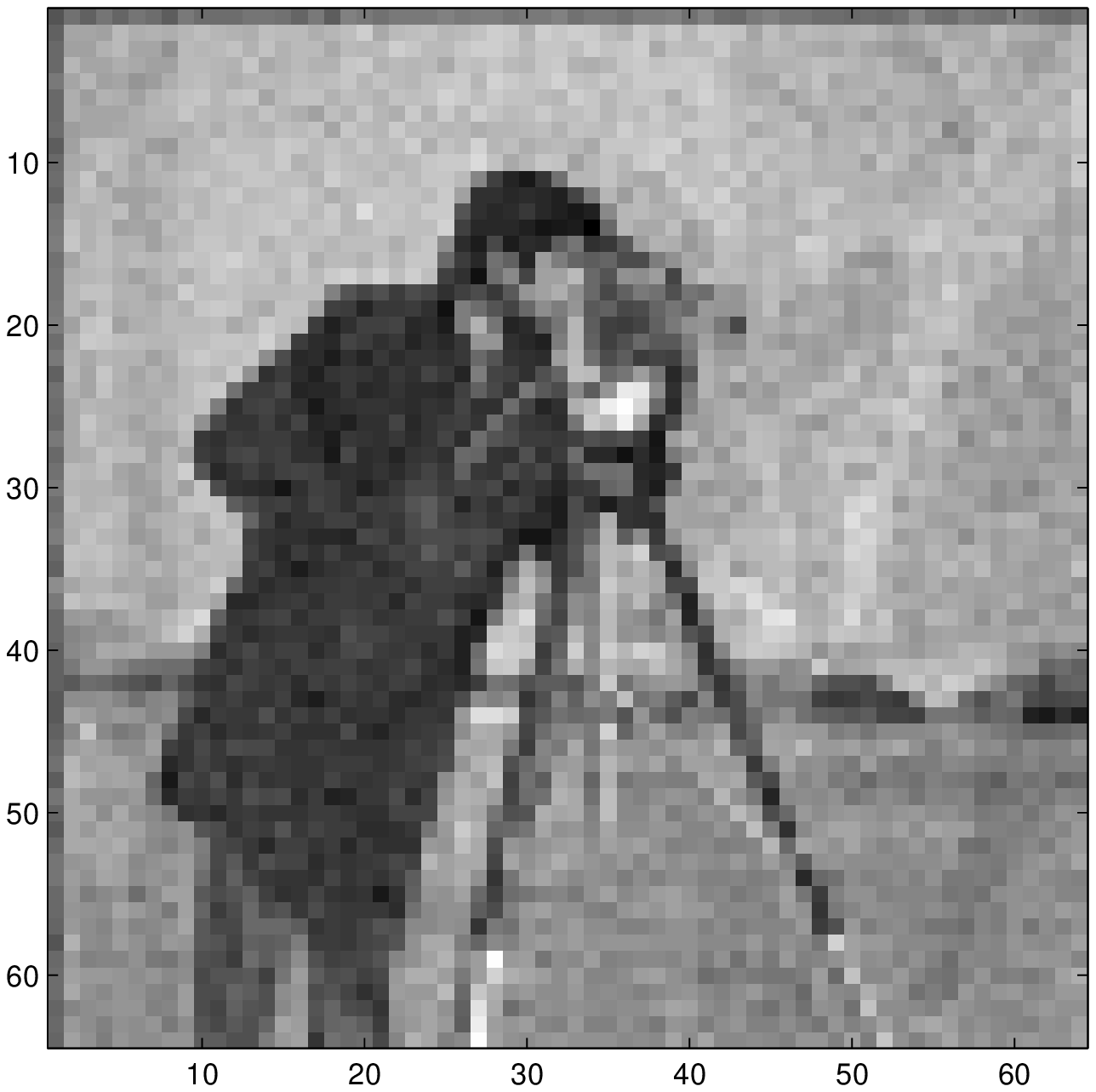} &
\includegraphics[width=45mm,height=45mm]{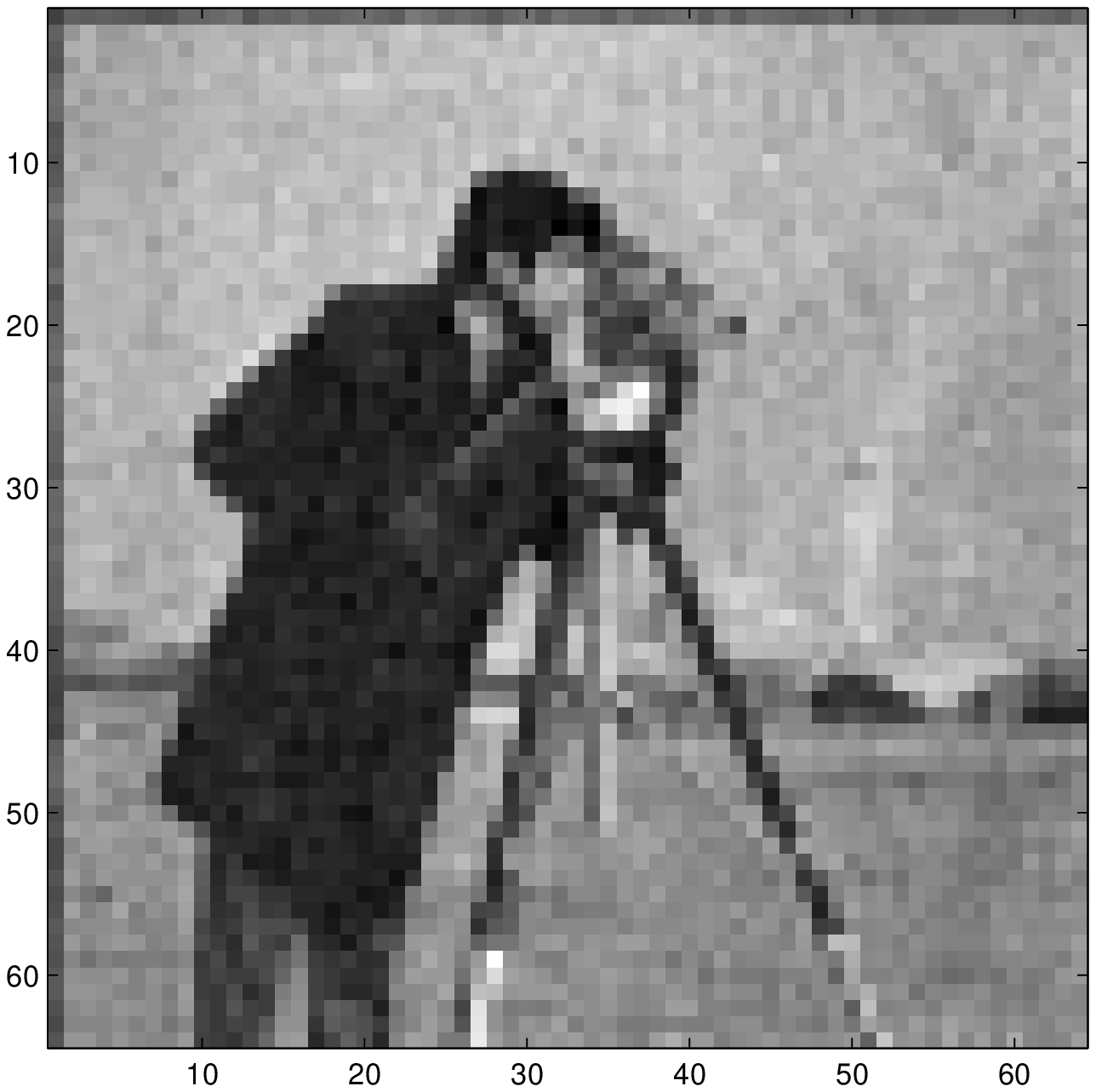} &
\includegraphics[width=45mm,height=45mm]{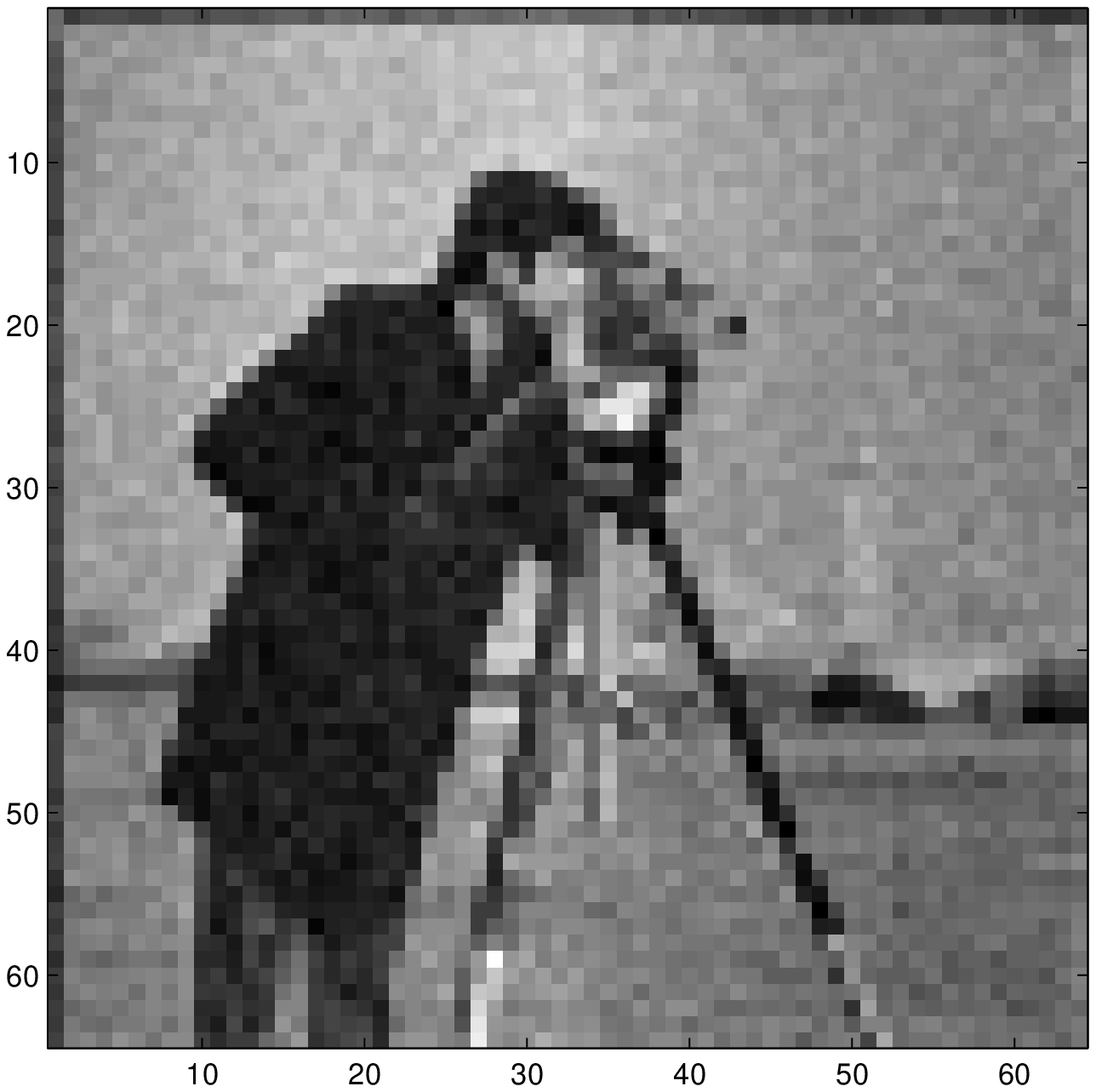} \\
PSNR = $0.1712$ & PSNR = $0.1049$ & PSNR = $0.1306$\\
\etabu
\caption{Estimated source images : (a) Sources assumed independent,
  (b) Accounting for local correlation in the sources, (c) 
Estimated sources obtained in the wavelet domain.}
\label{estimated_rect}
\efig

We may note that in this case which is an extremely favorable case the
three different methods give satisfactory results and it is not easy
to really distinguish between these three methods as it can also be
noted from the PSNR's of the reconstructed images compared to the
original images. We can, however, speculate that accounting for local
correlation of the image pixels outperforms the other two methods.

We have also considered a second case where
we have an equal number of measurements and sources (square case). The
original source images where mixed with the following matrix:

\[
A=\left[
\begin{array}{cc}
  0.9088 &  0.4928 \\
  0.4172 &  0.8702 
\end{array}
\right]
\] 
and the same type of noise was added to obtained the data with a SNR =
$30$dB shown in Figure \reff{mixed_squ}.

\bfig[!hbt]
\btabu{@{}c@{~}c@{}}
\includegraphics[width=45mm,height=45mm]{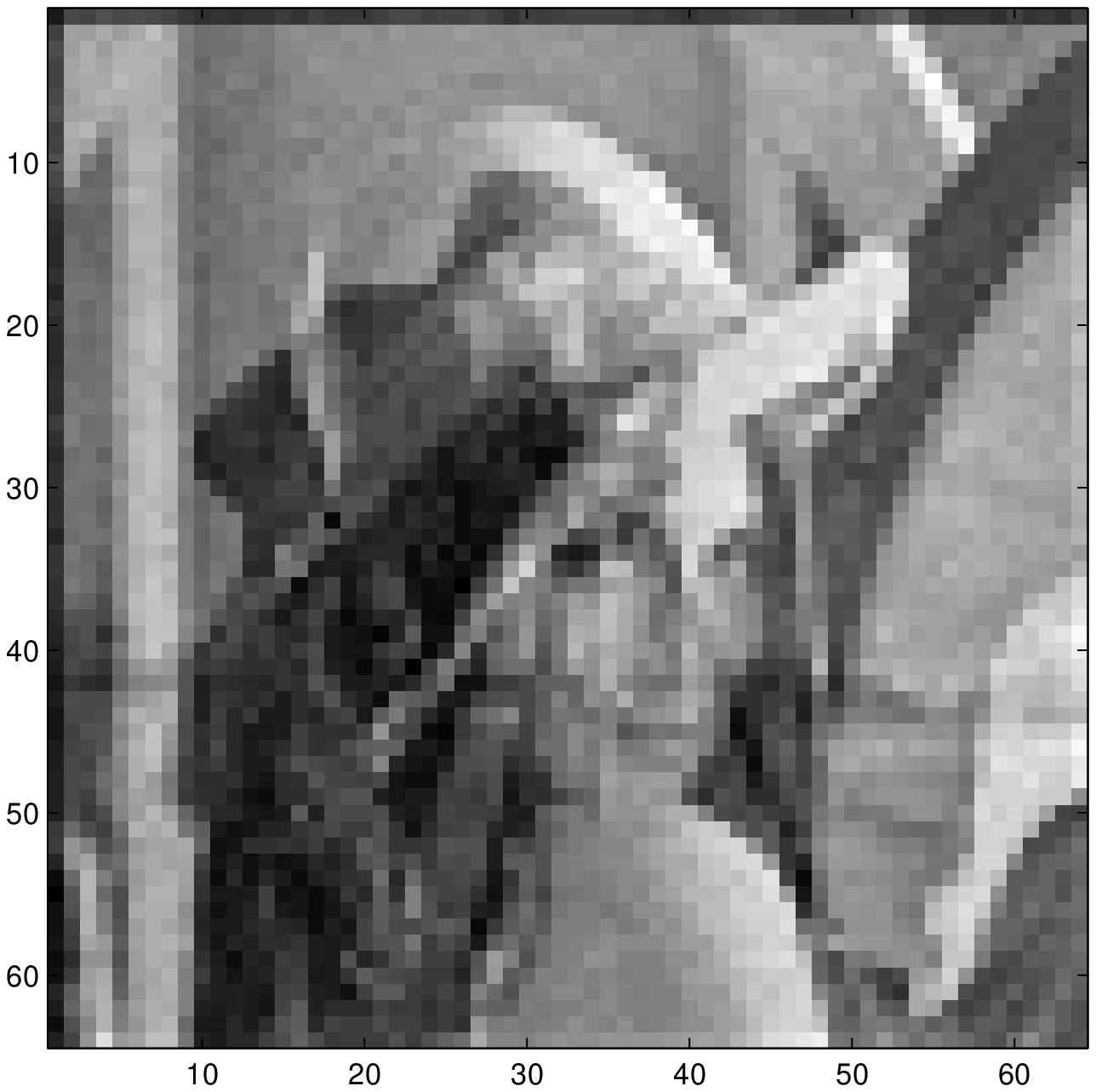} &
\includegraphics[width=45mm,height=45mm]{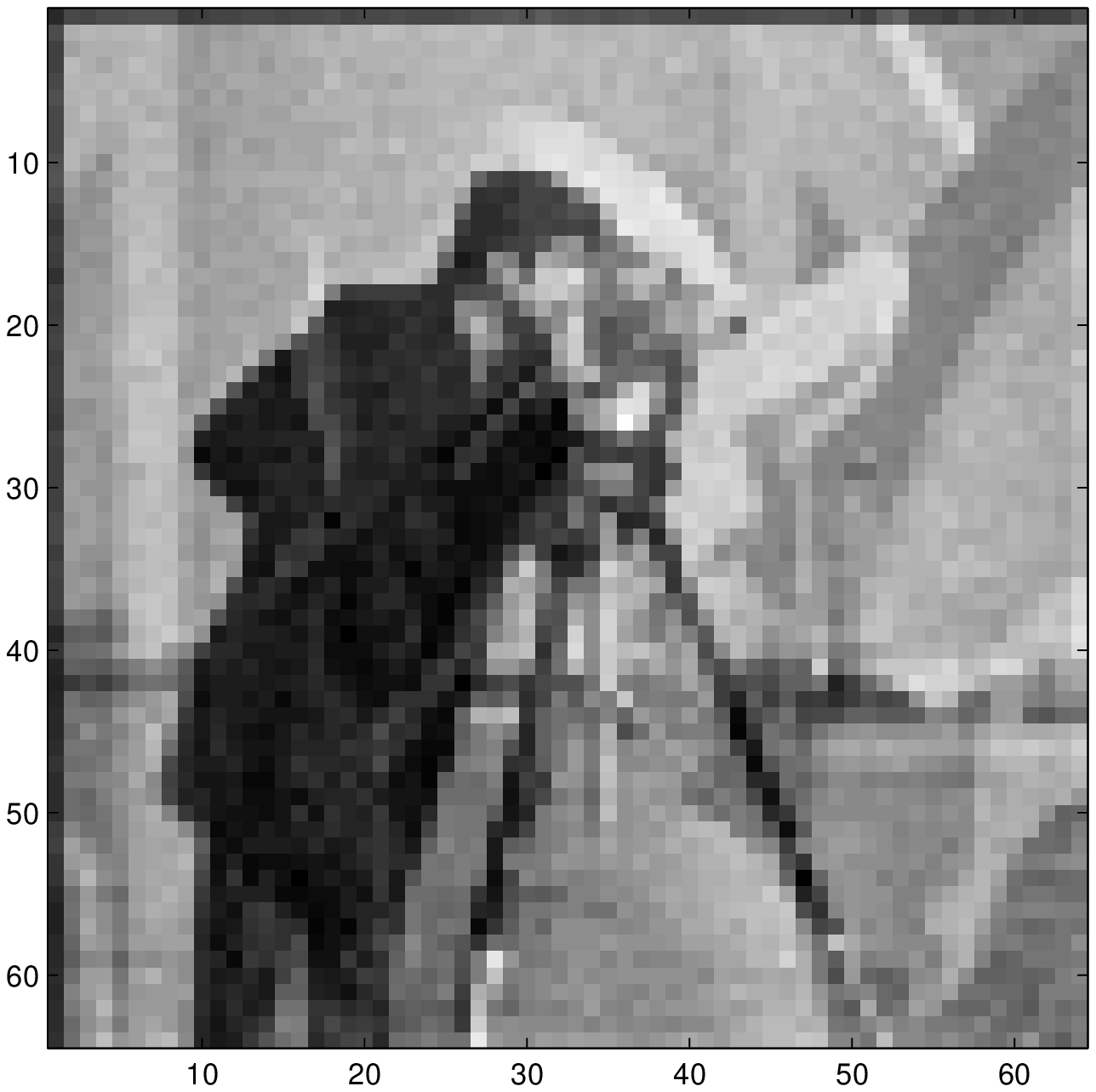}
\etabu
\caption{The mixed images for the case of a square mixing matrix}
\label{mixed_squ}
\efig

Figure \reff{estimated_squ} shows the reconstructed images by the three
methods of modeling the source images, i.e. Gaussian \iid,
Gauss-Markov on pixels and GE on their wavelet coefficients.

\bfig[!hbt]
\btabu{@{}c@{~}c@{~}c@{}}
(a) & (b) & (c) \\
\includegraphics[width=45mm,height=45mm]{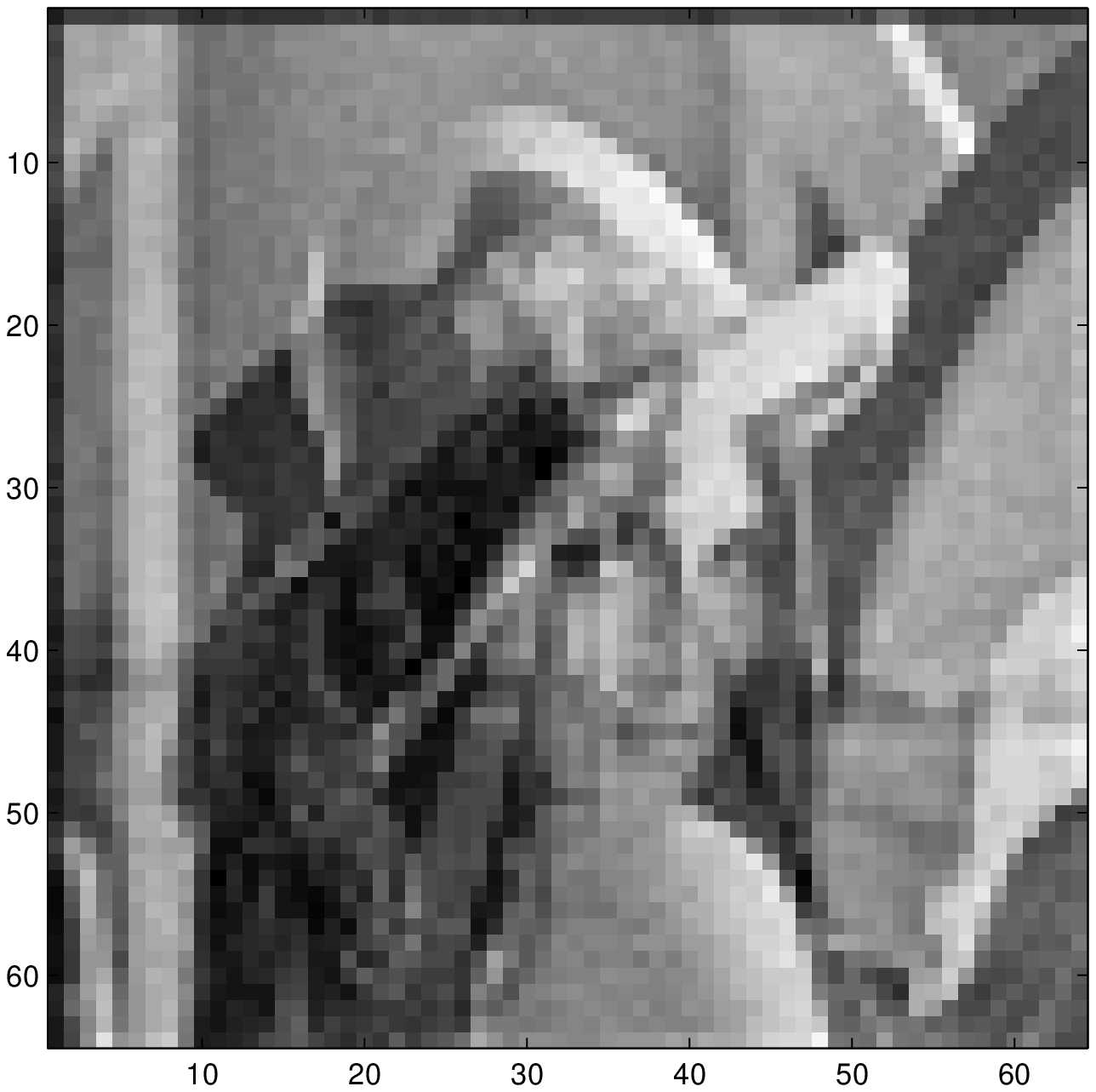} &
\includegraphics[width=45mm,height=45mm]{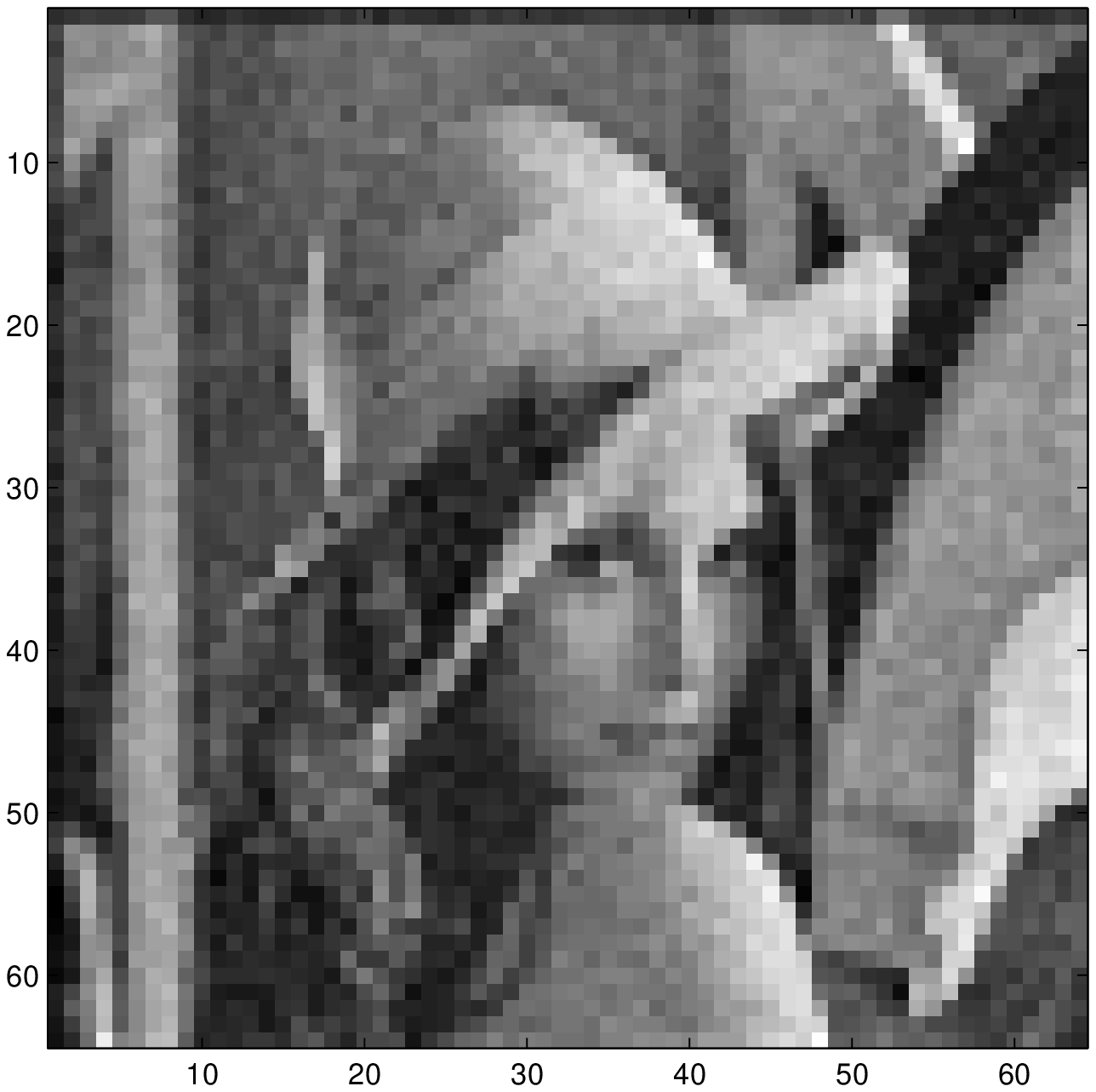} &
\includegraphics[width=45mm,height=45mm]{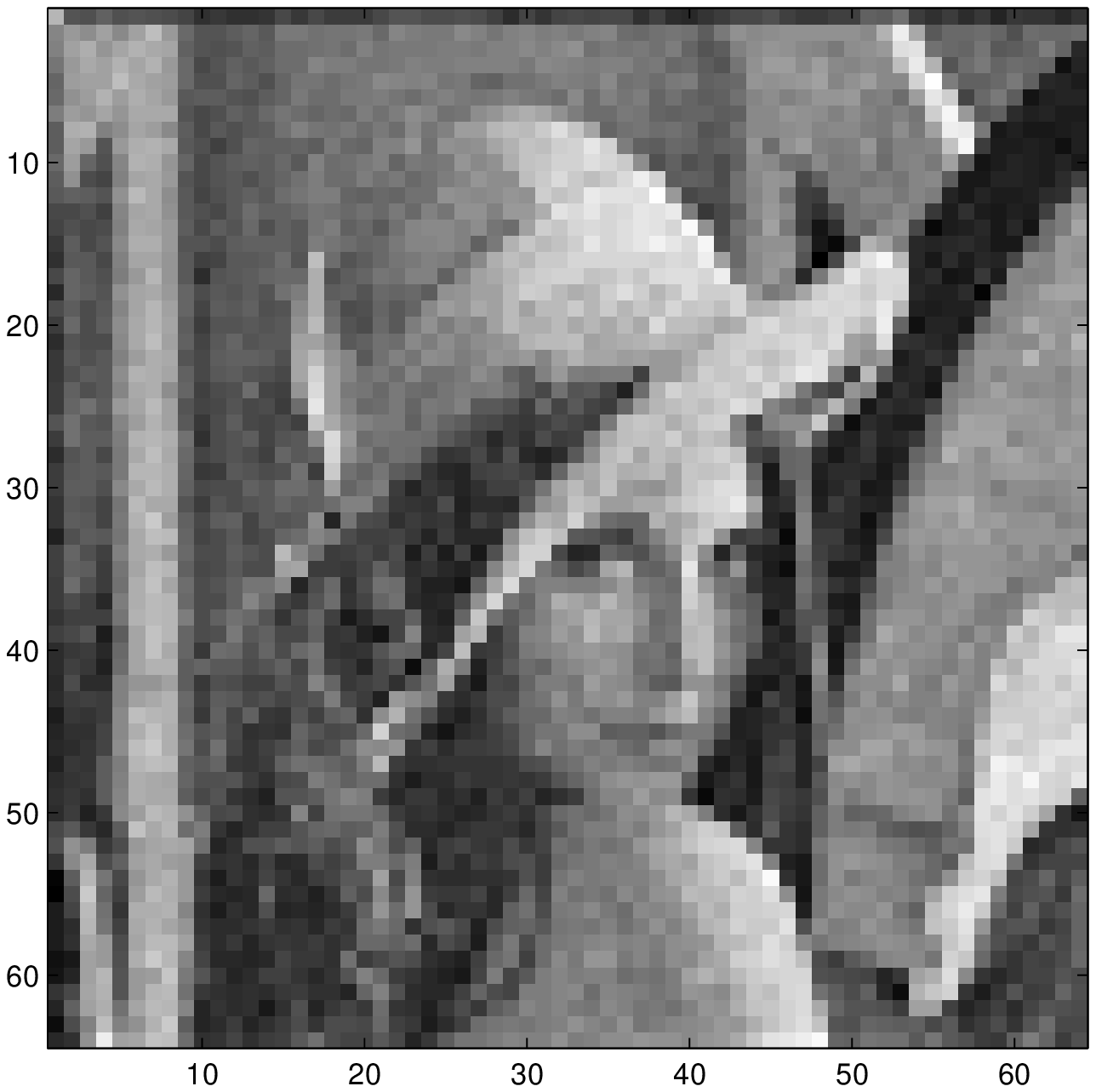} \\
PSNR = $0.2560$ & PSNR = $0.0740$ & PSNR = $0.1260$ \\
\includegraphics[width=45mm,height=45mm]{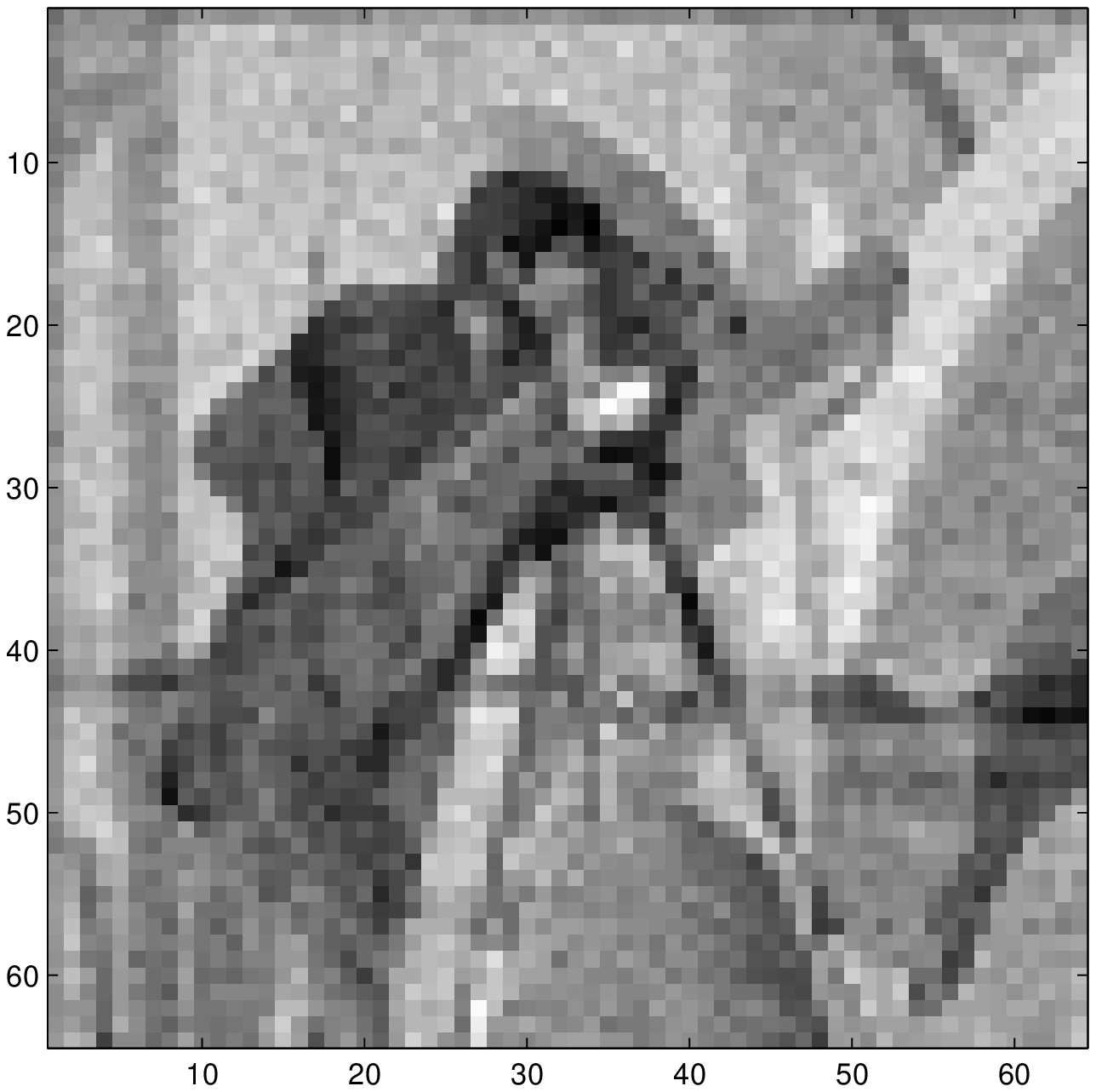} &
\includegraphics[width=45mm,height=45mm]{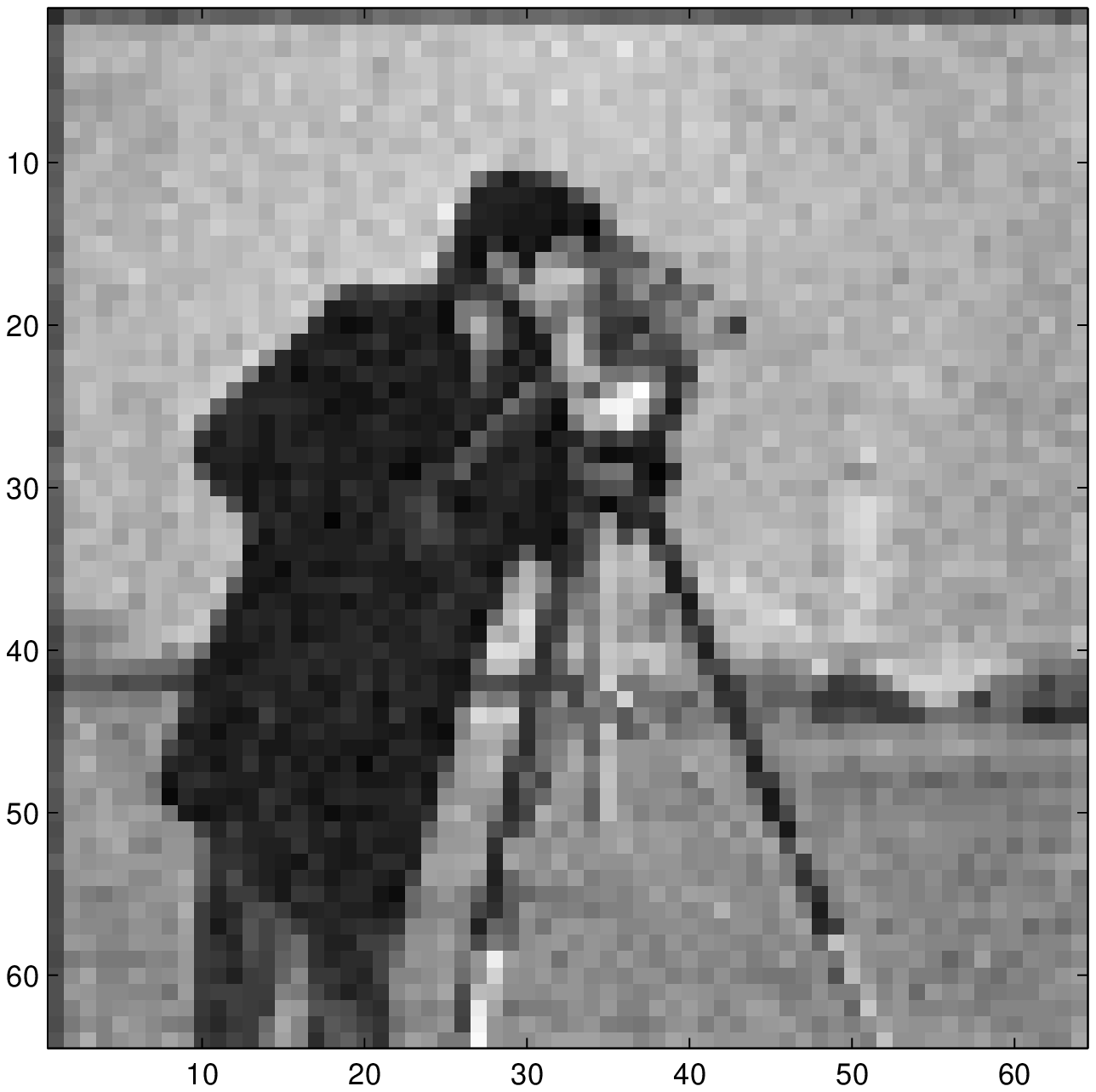} &
\includegraphics[width=45mm,height=45mm]{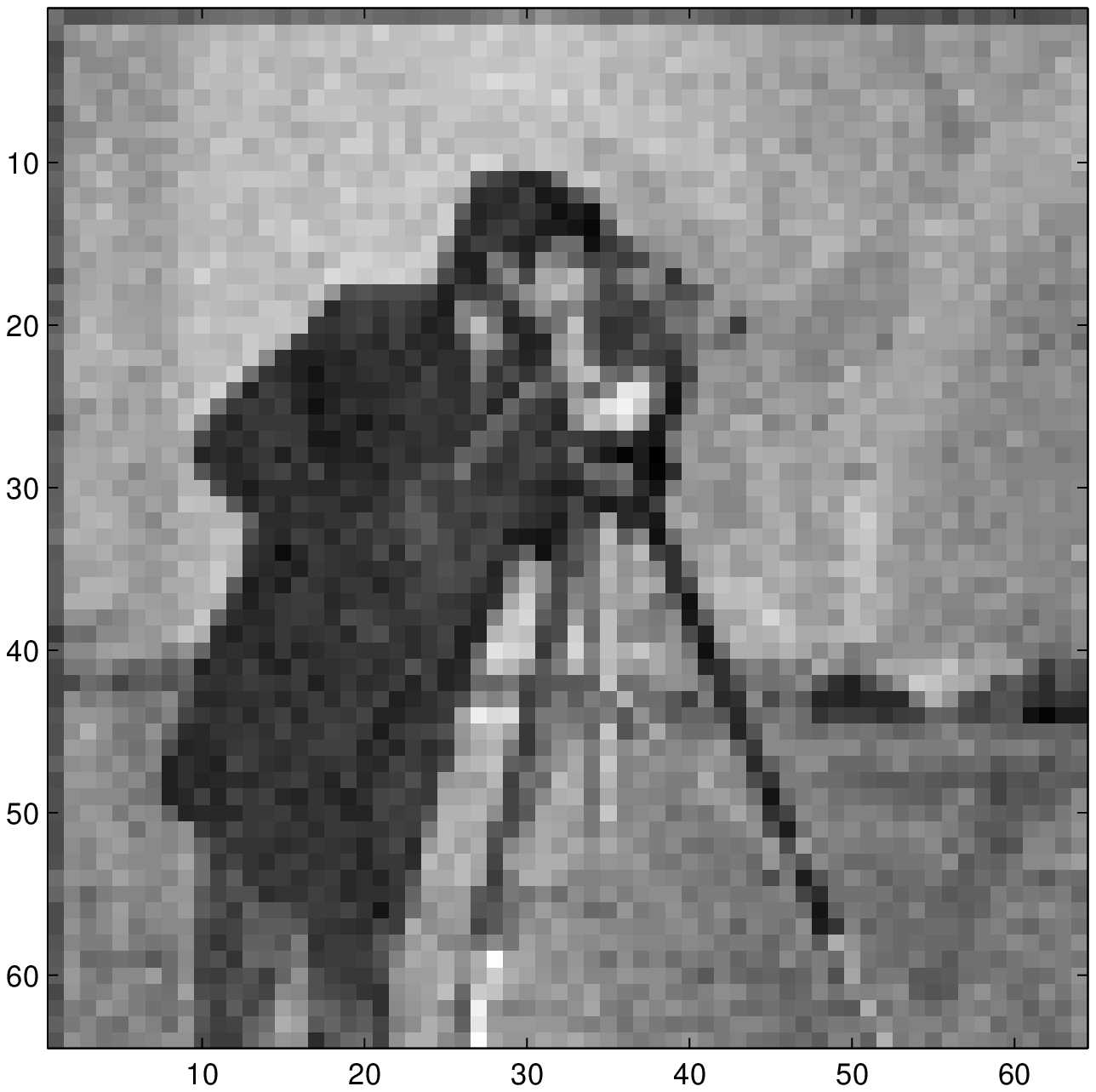} \\
PSNR = $0.3150$ & PSNR = $0.0998$ & PSNR = $0.1660$\\
\etabu
\caption{Estimated source images : (a) Sources assumed independent,
  (b) Accounting for local correlation in the sources, (c) 
Estimated sources obtained in the wavelet domain.}
\label{estimated_squ}
\efig

We should point out that we have used the following values for the initialization of the algorithm:
\[
\Ab^{(0)}=\left[
\begin{array}{cc}
1.0 & 0.0 \\
0.0 & 1.0
\end{array}
\right], \quad 
\sigma^{2^{(0)}}=\sigma_1^{2^{(0)}}=\sigma_2^{2^{(0)}}=1 
\longrightarrow 
\left \{
\begin{array}{l}
\ale^{(0)}=\sqrt{2}~,~\beta =2 \\
\als^{(0)}=(2)^{1/\beta _s}~,~\beta _s=1.9 
\end{array}
\right.
\]

The final estimated values obtained by averaging the last 10\% samples 
after 5000 iterations are the following:
\begin{displaymath}
\left.
\begin{array}{cc}
\wh{\Ab}=\left[
\begin{array}{cc}
 0.8604  & 0.4681 \\
 0.5096  & 0.8837   
\end{array}
\right]
&
\begin{array}{l}
 \hat{\alpha}_{\epsilon_1} = 24.0966,~ \hat{\alpha}_{\epsilon_2} = 24.2096 \\
 \hat{\alpha}_{s_1} = 91.4272,~ \hat{\alpha}_{s_2} = 83.5939 
\end{array}
\end{array}
\right.
\end{displaymath}

We may also note that the estimated values of $\alpha_{\epsilon_1}$,
$\alpha_{\epsilon_2}$, $\alpha_{s_1}$ and $\alpha_{s_2}$
directly from the original images are:
\begin{center}
\begin{tabular}{l}
$ \alpha_{\epsilon_1} =7.6457,~ \alpha_{\epsilon_2} = 7.2784 $ \\
$ \alpha_{s_1} = 96.5342,~ \alpha_{s_2} = 107.9316 $
\end{tabular}
\end{center}

We notice that neither the noise variances nor the variance of the
second image (the cameraman) were well estimated. We clearly notice that in Figure
\reff{estimated_squ}. However, the separation of the images in the
wavelet domain outperforms the separation applied directly to the
images assuming sources to be independent and this is due to the
decorrelation property of the wavelet transform. In fact, the wavelet
transform nearly decorrelates a signal, thus assuming independent
wavelet coefficients is more realistic than assuming independent signal samples.

Figure \reff{rate} shows the rate of acceptance of the generated
samples from the Gaussian to approximate the posterior law of the
wavelet coefficients for $\beta_s=1.9$. 

\bfig[!htb]
\includegraphics[width=80mm,height=35mm]{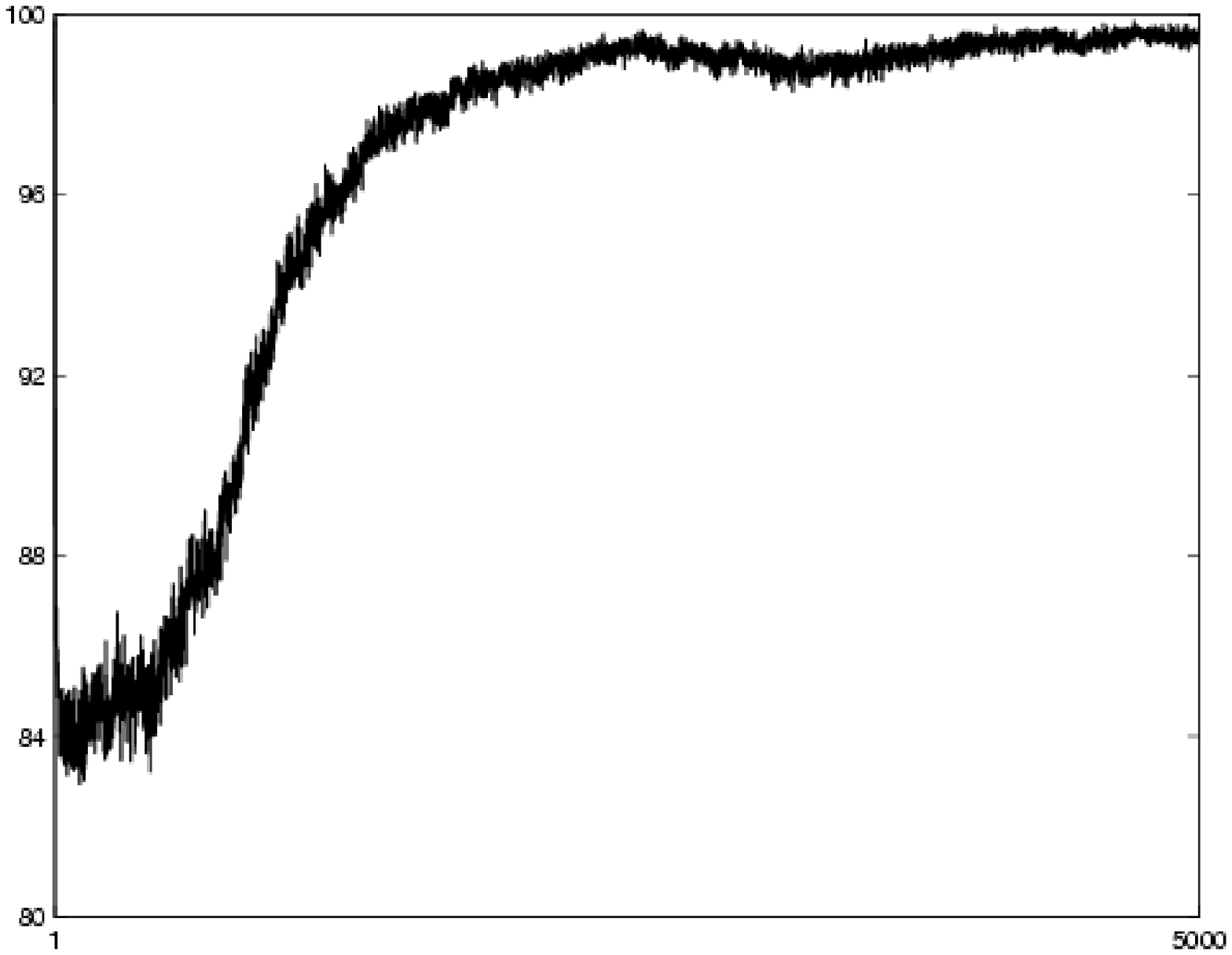}
\caption{Rate of acceptance of the samples for the wavelet
  coefficients along the iterations}
\label{rate}
\efig

We also noticed that this rate of acceptance is a function of the
parameter $\beta_s$:
$$ \rho \searrow 0~~~~ \textrm{as}~~~~\beta_s \searrow 1 $$
and 
$$ \rho \nearrow 1~~~~ \textrm{as}~~~~\beta_s \nearrow 2 $$

Figure \reff{conv_a} shows
the convergence of the elements of the matrix $\Ab$ and Figure
\reff{conv_alpha} shows the convergence of the hyperparameters. 

\bfig[!htb]
\btabu{@{}c@{~}c@{}}
\includegraphics[width=65mm,height=30mm]{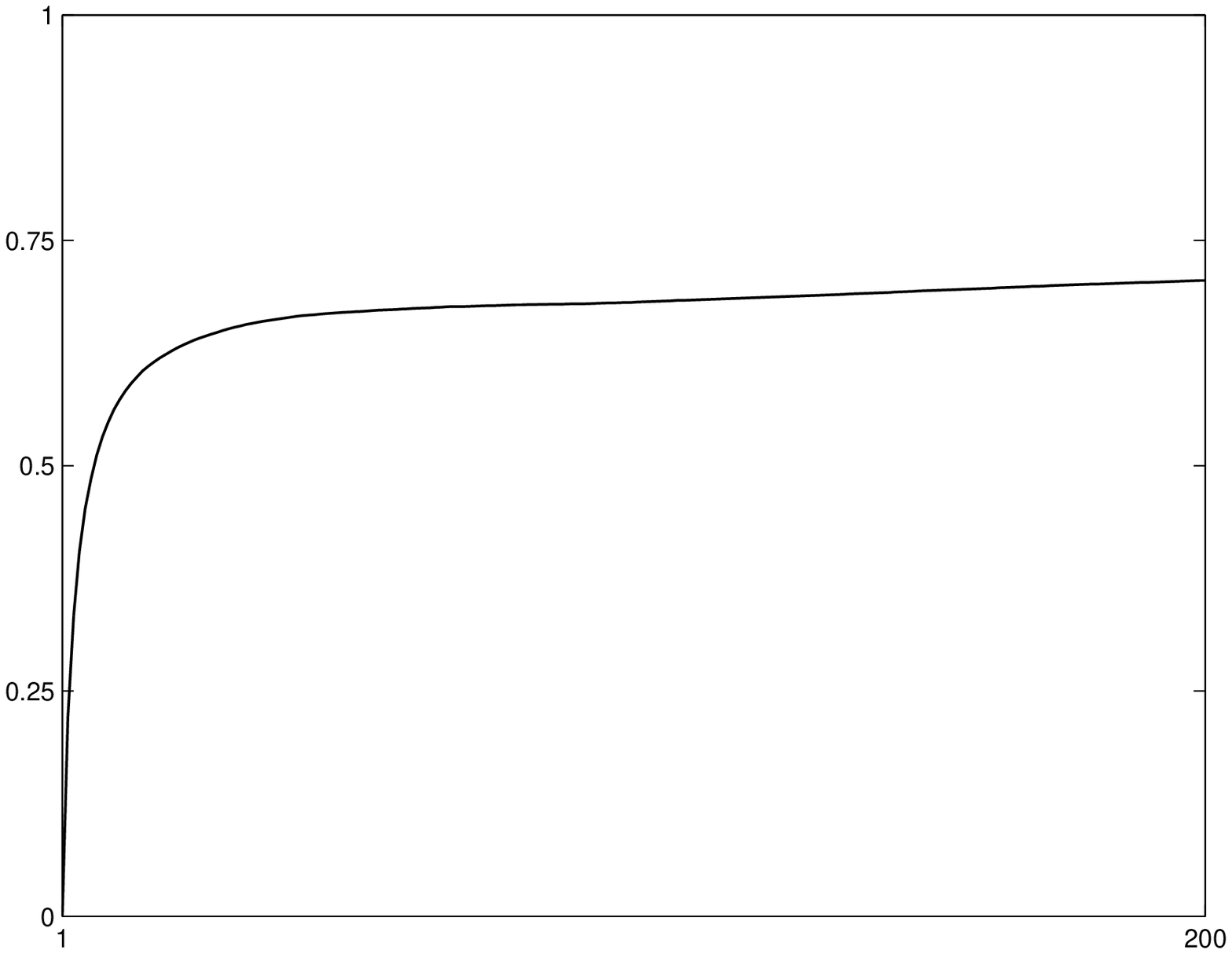} &
\includegraphics[width=65mm,height=30mm]{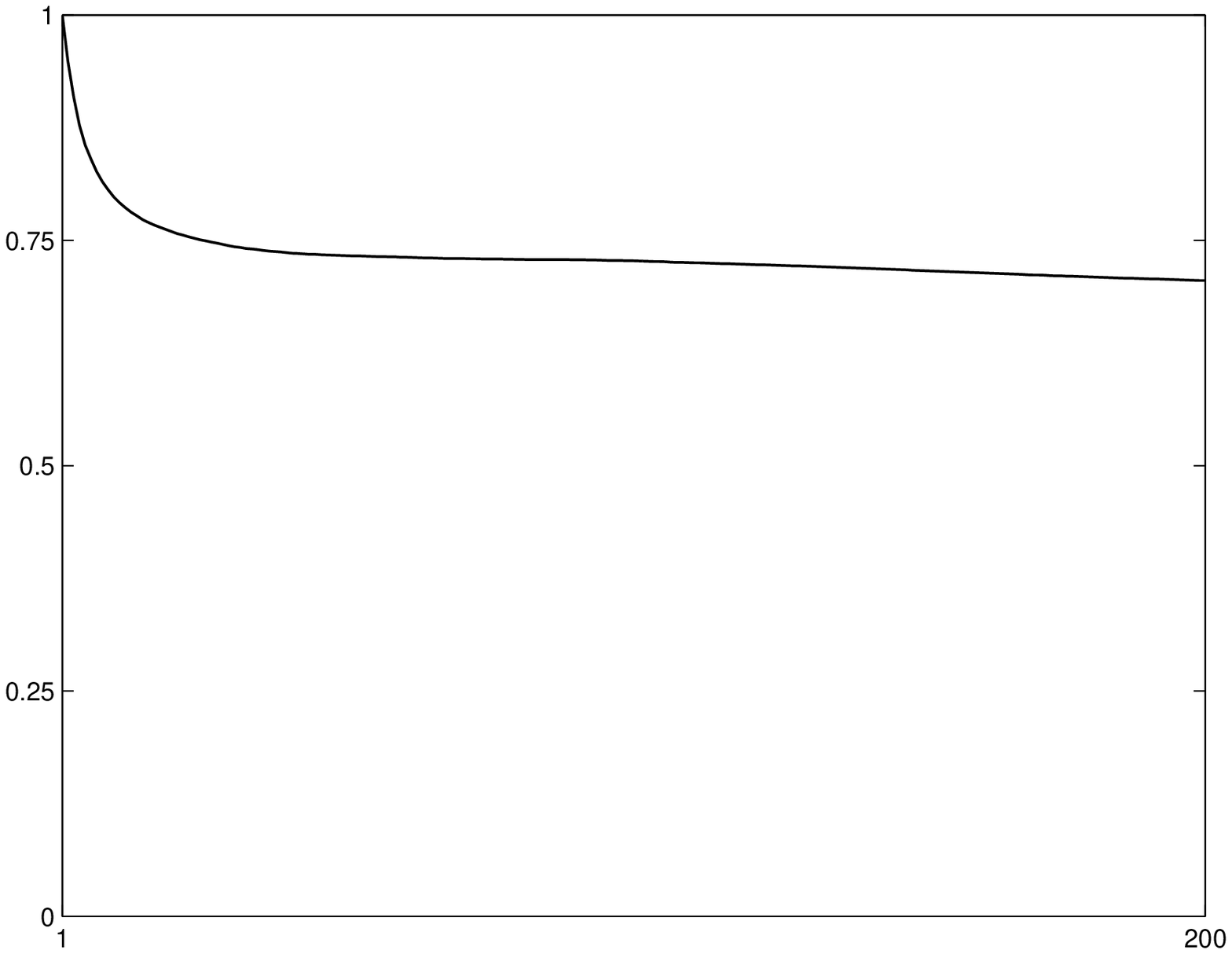} \\
$a_{11}$ & $a_{12}$ \\
\includegraphics[width=65mm,height=30mm]{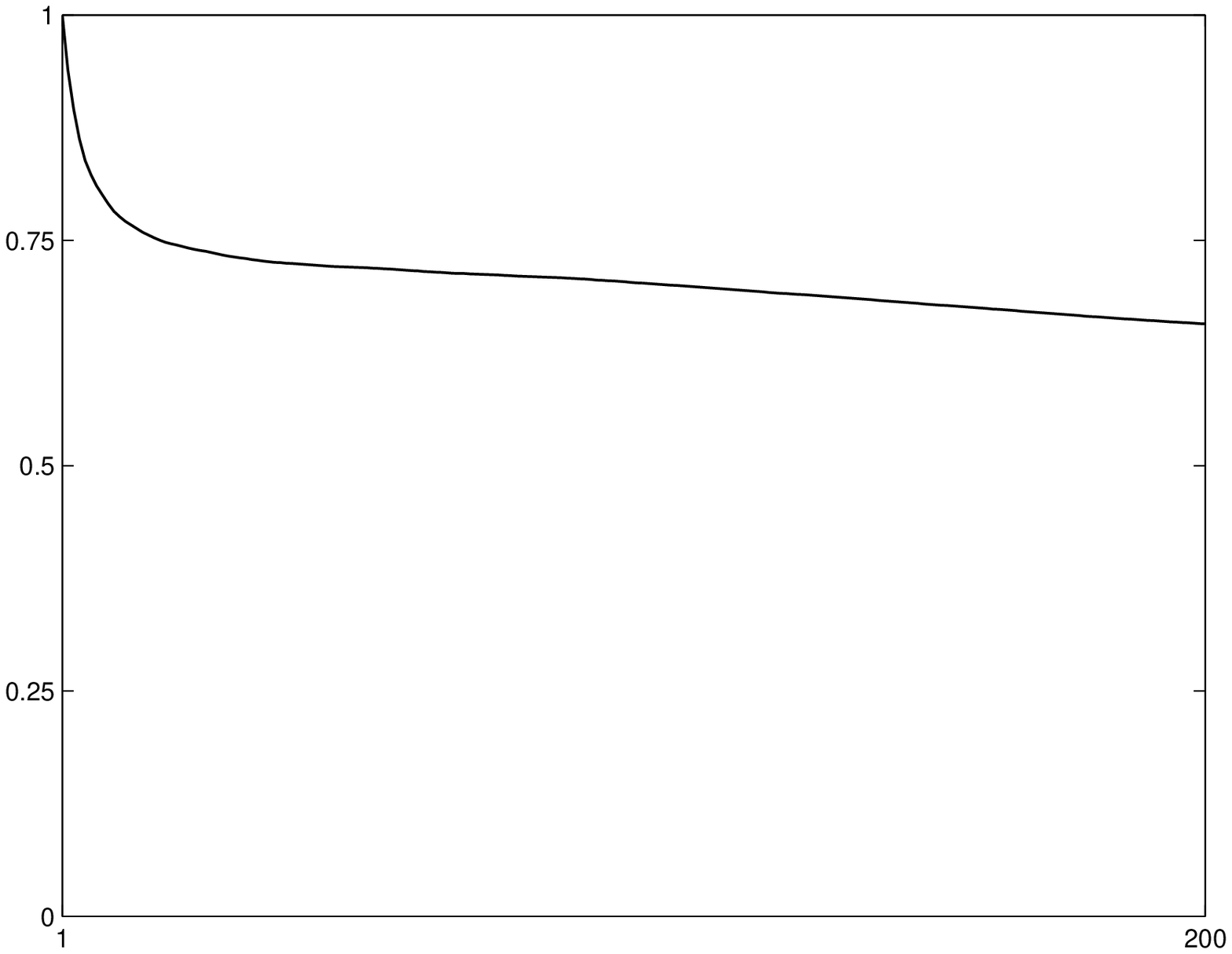} &
\includegraphics[width=65mm,height=30mm]{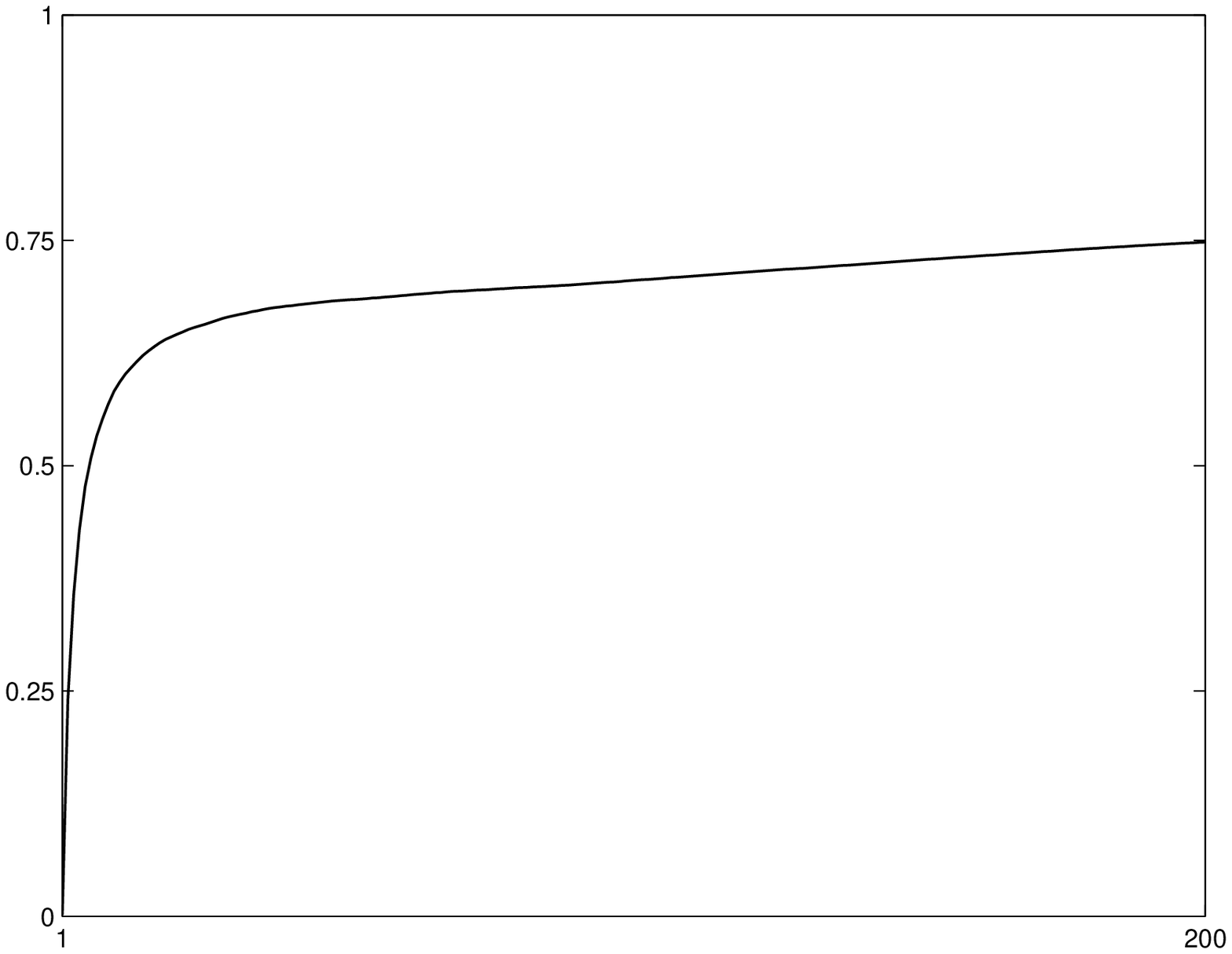} \\
$a_{21}$ & $a_{22}$ \\
\etabu
\caption{Convergence of the elements of $\Ab$ during the first $200$ iterations}
\label{conv_a}
\efig

\bfig[!htb]
\btabu{@{}c@{~}c@{}}
\includegraphics[width=65mm,height=35mm]{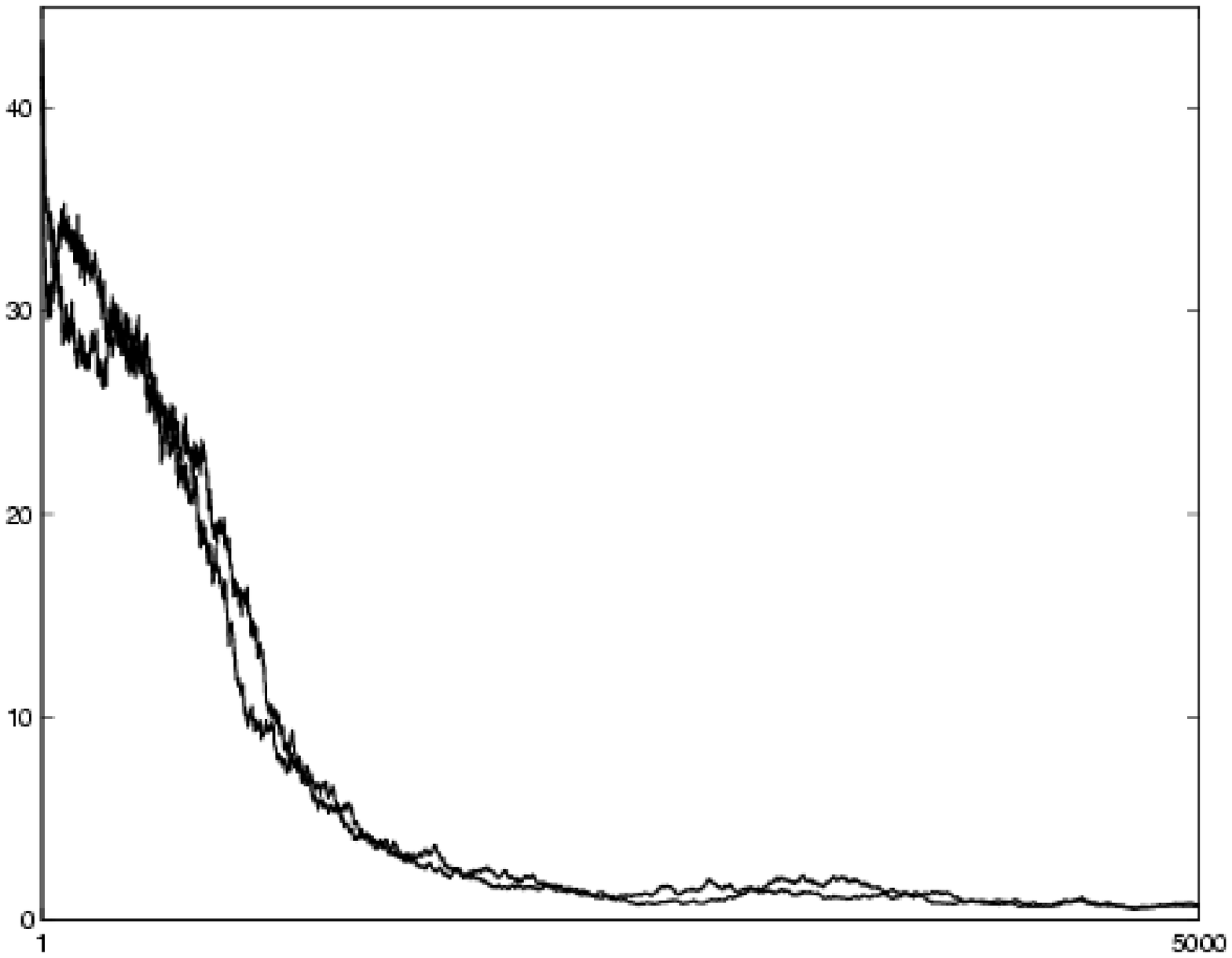} & 
\includegraphics[width=65mm,height=35mm]{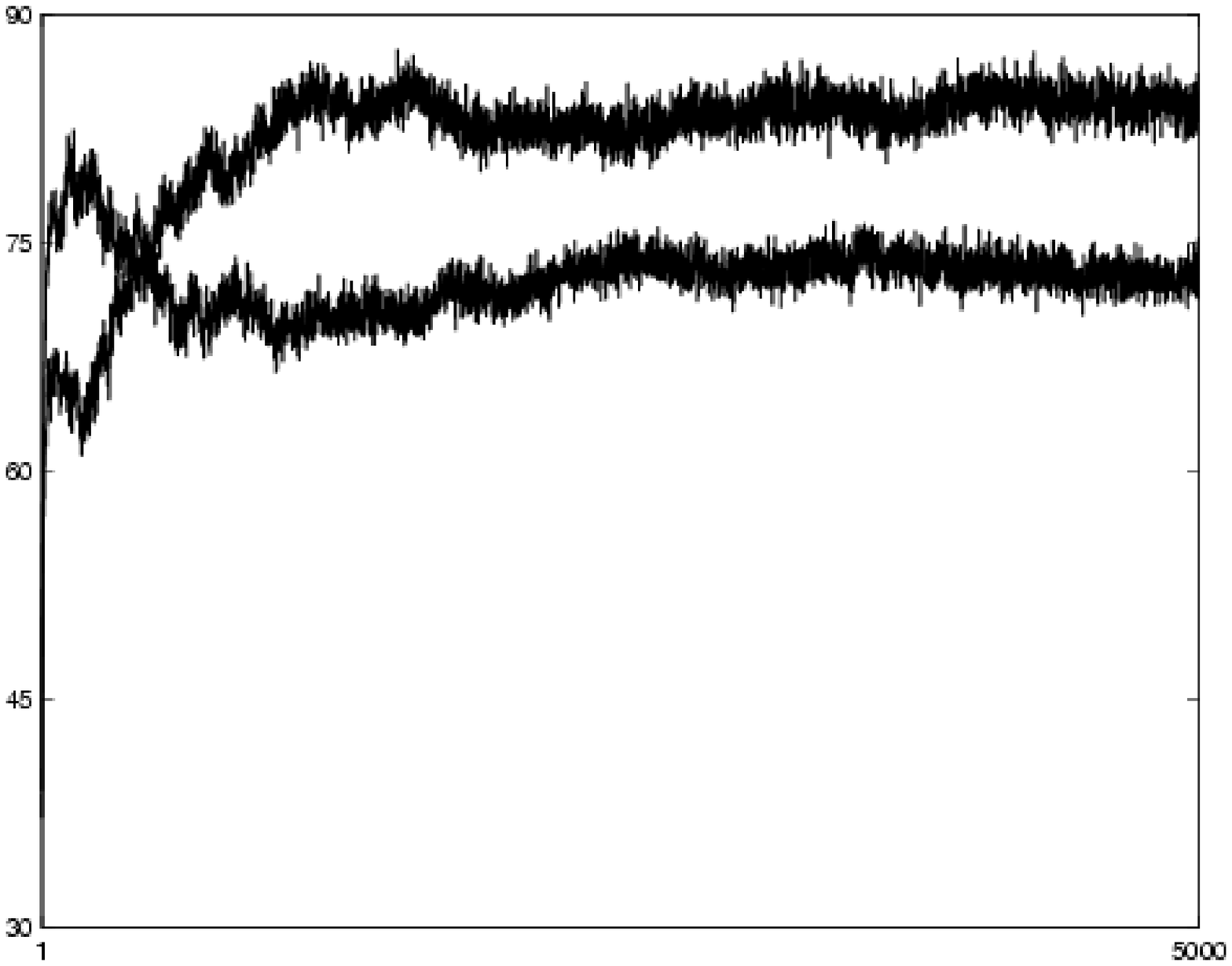} \\
$\alpha_\epsilon$ & $\alpha_s$
\etabu
\caption{Convergence of the hyperparameters $\thetab$: 
Left: $\alpha_{\epsilon 1}$ and $\alpha_{\epsilon 2}$ \quad Right:
$\alpha_{s1}$ and $\alpha_{s2}$.}
\label{conv_alpha}
\efig
\noindent
Figure \reff{hist} shows the histograms of the original and
estimated images while Figure \reff{histw} shows the histograms of the wavelet coefficients
of the original images superposed with the Exponential pdf with parameter
$\alpha$ estimated with the algorithm.

\bfig[!htb]
\btabu{@{}c@{~}c@{}}
\includegraphics[width=65mm,height=35mm]{ho1} &
\includegraphics[width=65mm,height=35mm]{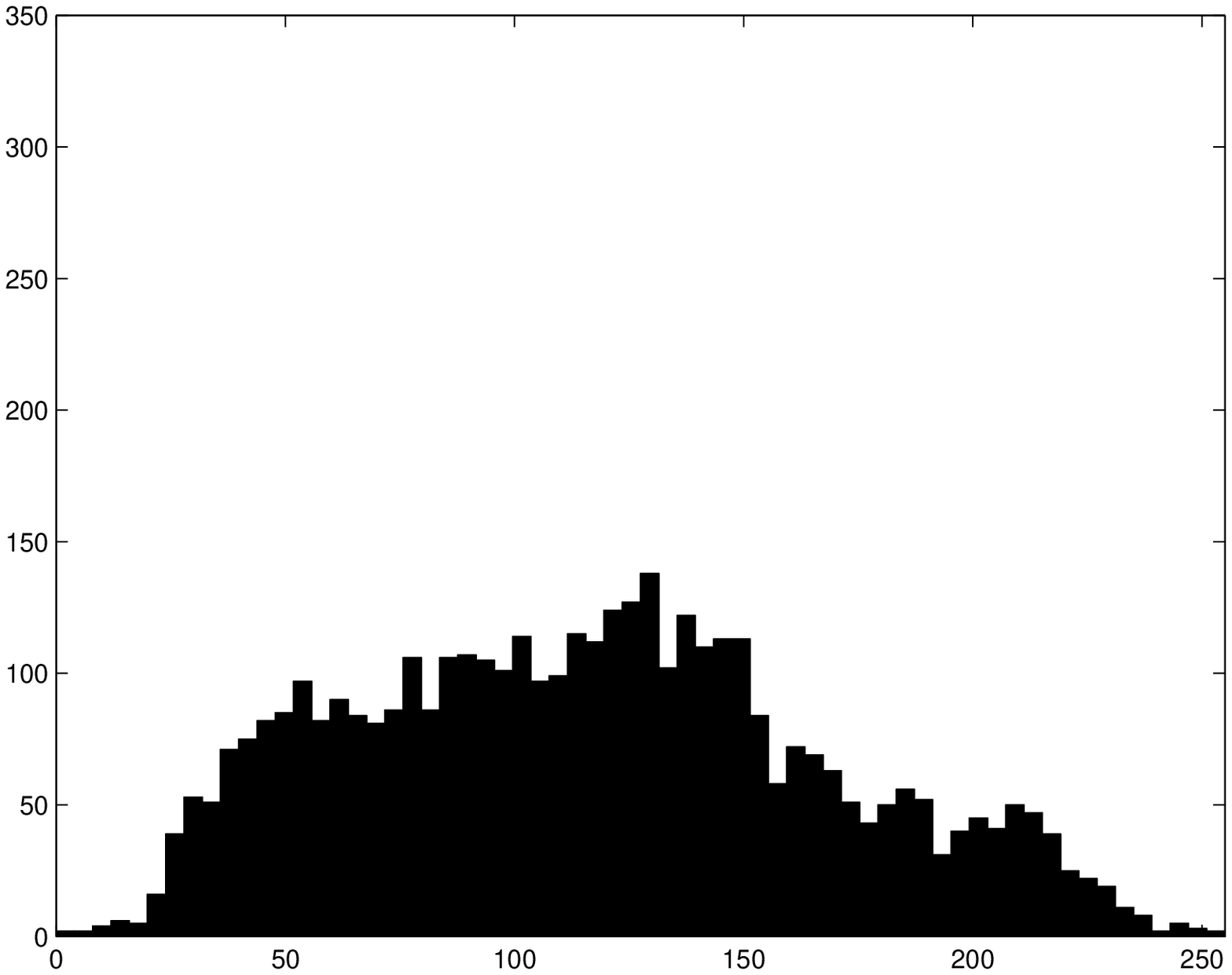} \\
\includegraphics[width=65mm,height=35mm]{ho2} & 
\includegraphics[width=65mm,height=35mm]{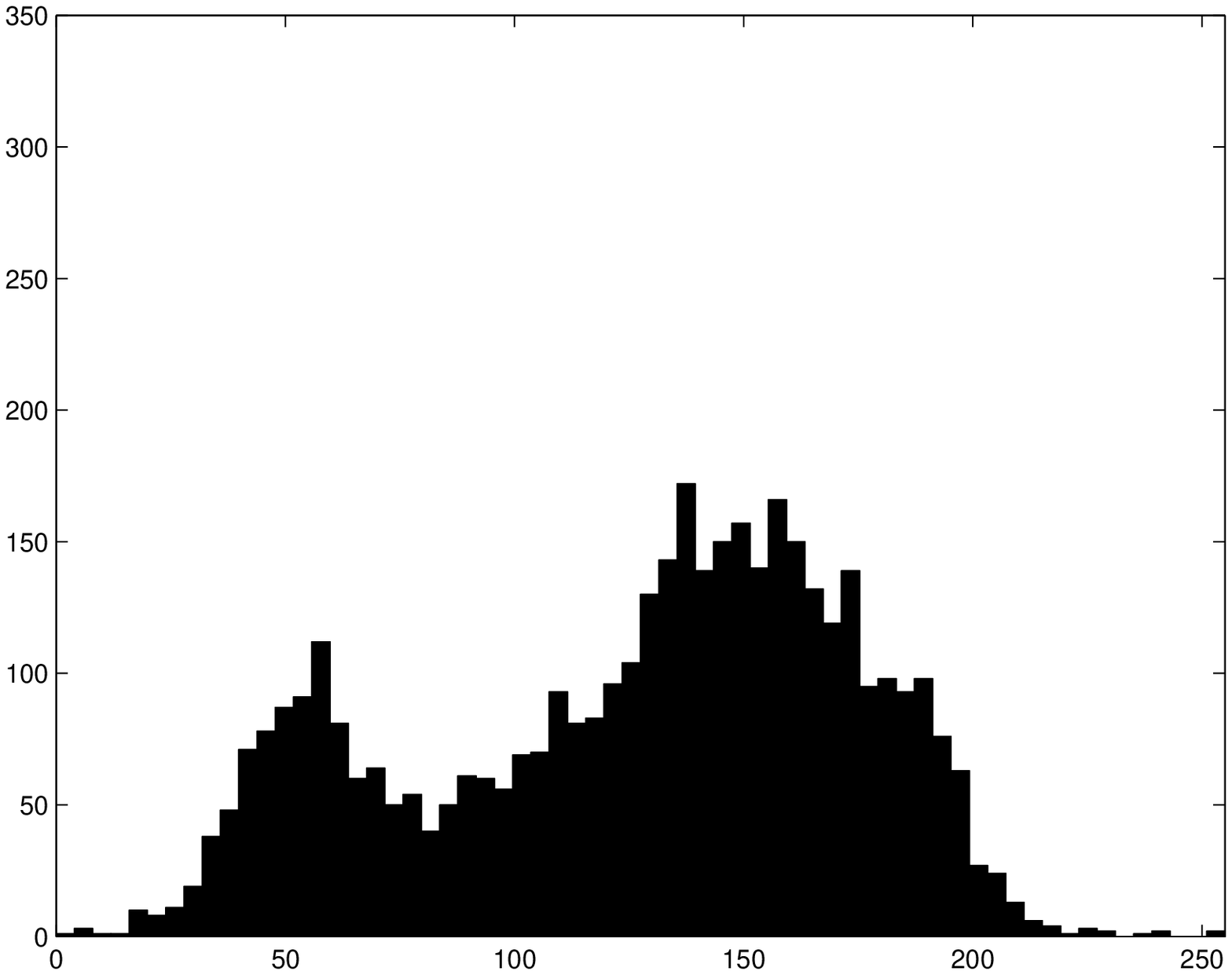} \\
(a) & (b) \\
\etabu
\caption{The histogram of: (a) Original source images, (b) The
  estimated images (top: Lena image, bottom: The cameraman image)}
\label{hist}
\efig

\bfig[!htb]
\btabu{@{}c@{~}c@{}}
\includegraphics[width=65mm,height=35mm]{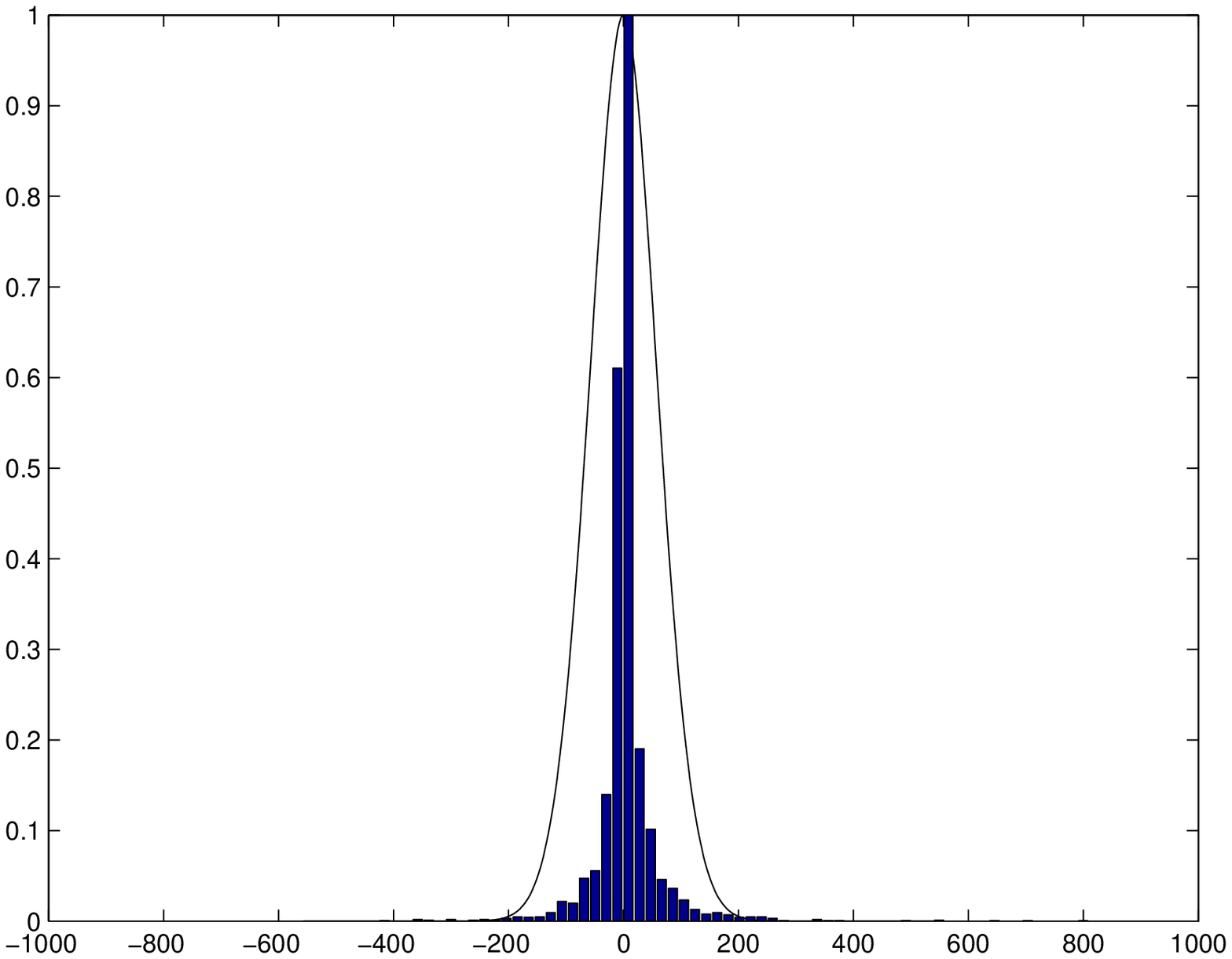} &
\includegraphics[width=65mm,height=35mm]{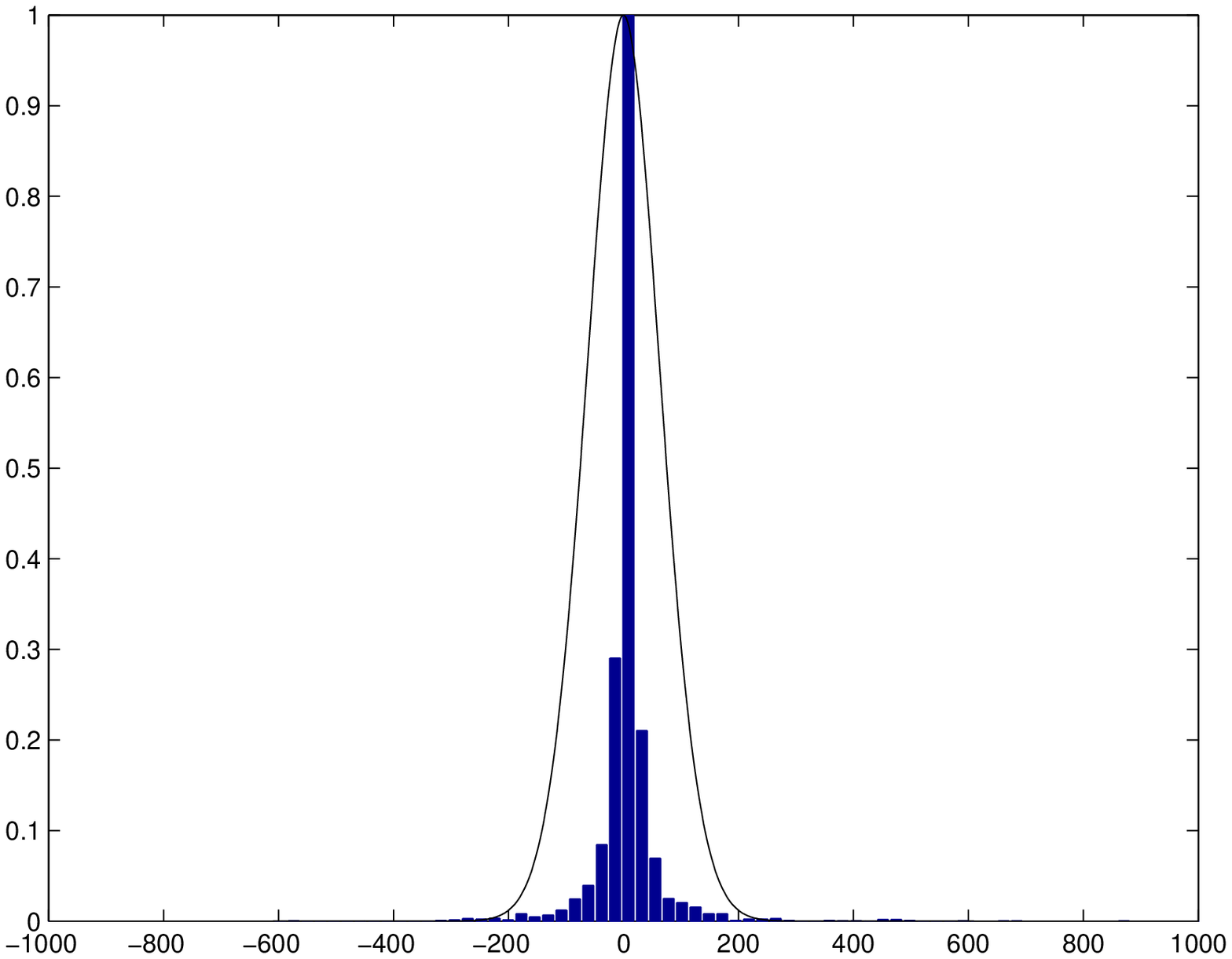} \\
(a) & (b) \\
\etabu
\caption{The histogram of the wavelet coefficients of the 
original source images superposed with the pdf of the estimated
images of: (a) Lena image, (b) The cameraman image}
\label{histw}
\efig

\section{Conclusions and Perspectives}
\label{Conclusions}
In this contribution we proposed an approach to jointly estimate the
mixing matrix and the original source images. We transported the
problem to the wavelet domain using a Bayesian approach where the wavelet
coefficients of real world images are naturally modeled by
generalized exponential distributions. Independence of the wavelet
coefficients of signals is more realistic than the independence of the
signals themselves.

In a first step, we assumed all the wavelet coefficients to be independent 
and identically distributed and follow a GE pdf with a fixed value for
its parameter $\beta_s$ while its second parameter is estimated during
the iterations. Even if this gives satisfactory results, it will
be better to estimate $\beta_s$ too during the iterations.

A second point is that the choice of a Gaussian trial pdf is good when
$\beta_s$ is not far from $2$, but it seems that this choice is no
more efficient when $\beta_s$ approaches $1$.

Finally, since the wavelet coefficients of real world signals (images)
tend to propagate through scales, a future work is to put a Markovian
model on the wavelet coefficients taking into account inter-scale
correlation of the coefficients.

\bibliographystyle{ieeetr}     % Style de bibliographie
\bibliography{revuedef,revueabr,bibfr,baseAJ,baseKZ,gpipubli,ss,cardozo,mackay,ica99,mybib,mybib2,biben}

\end{document}